\documentclass[aps,pre,twocolumn,floatfix]{revtex4}

\usepackage{bm}
\usepackage{graphicx}
\usepackage{amssymb,amsfonts,amsmath}
\usepackage{epsf}
\usepackage{color}
\usepackage{hyperref}
\usepackage{subfigure}
\usepackage{epstopdf}
\usepackage{soul}
\usepackage{mathrsfs}

\usepackage{upgreek}

\newcommand{\be}{\begin{equation}}
\newcommand{\ee}{\end{equation}}
\newcommand{\bea}{\begin{eqnarray}}
\newcommand{\eea}{\end{eqnarray}}

\begin{document}

\title{\bf Vacancy diffusion and the hydrodynamics of crystals}

\author{Jo\"el Mabillard}
\email{Joel.Mabillard@ulb.be; \vfill\break ORCID: 0000-0001-6810-3709.}
\author{Pierre Gaspard}
\email{Gaspard.Pierre@ulb.be; \vfill\break ORCID: 0000-0003-3804-2110.}
\affiliation{Center for Nonlinear Phenomena and Complex Systems, Universit{\'e} Libre de Bruxelles (U.L.B.), Code Postal 231, Campus Plaine, B-1050 Brussels, Belgium}

\begin{abstract}
The hydrodynamics of crystals with vacancies is developed on the basis of local-equilibrium thermodynamics, where the chemical potential of vacancies plays a key role together with a constraint relating the concentration of vacancies to the density of mass and the strain tensor.  The microscopic foundations are established, leading to Green-Kubo and Einstein-Helfand formulas for the transport coefficients, including the vacancy conductivities and the coefficients of vacancy thermodiffusion.  As a consequence of having introduced the chemical potential of vacancies, a relationship is obtained between the conductivities and the Fickian diffusion coefficients for the vacancies.  The macroscopic equations are linearized around equilibrium to deduce the dispersion relations of the eight hydrodynamic modes.  The theoretical predictions are confirmed by numerical simulations of the hard-sphere crystal with vacancies.  The study explicitly shows that the eighth hydrodynamic mode of nonperfect monatomic crystals is indeed a mode of vacancy diffusion.

\end{abstract}

\maketitle

\section{Introduction}
Crystals form periodic spatial structures in contrast to fluids, which have uniform properties at equilibrium.  In the crystalline phase, the continuous symmetries under three-dimensional spatial translations are thus broken.  As a consequence of the Nambu-Goldstone mechanism~\cite{N60,G61}, there exist three slow modes in addition to the five modes associated with the five fundamentally conserved quantities, which are mass, energy, and linear momentum, for the macrodynamics of one-component crystals \cite{F75}.  Therefore, monatomic crystals have eight hydrodynamic modes, including the six longitudinal and transverse sound modes, the heat mode, and an extra mode identified as the vacancy diffusion mode in the early seventies \cite{MPP72,FC76}.  In perfect crystals, i.e., crystals without vacancies, the eighth mode is absent and there exist only seven modes, which have been extensively studied \cite{KDEP90,MG24a,MG24b,MG24c,PBHB24}.  Besides, there is a vast literature on vacancies, which are point defects, and their diffusion.  On the one hand, much work has been devoted to the equilibrium properties of vacancies \cite{HL64,AL87,PF01,OGHLRS10,FGHNKJV14,L20}, in particular, in crystalline solids under stress \cite{LC78,LC85,CVJ18}.  On the other hand, the diffusion of vacancies has been thoroughly investigated in crystals under nonequilibrium conditions \cite{HL64,AL87,CL83,KBR89}.  In these studies, the chemical potential of vacancies plays an essential role, not only to obtain the equilibrium concentration of vacancies, but also to drive their diffusion by the gradient of their chemical potential.  However, the importance of the chemical potential of vacancies does not seem to have been fully appreciated for understanding how Fick's law of vacancy diffusion may arise in the hydrodynamics of crystals.  The chemical potential of vacancies is briefly mentioned in Ref.~\cite{FC76}, but without explicitly explaining how a genuine Fickian diffusion coefficient may arise for vacancies within the framework of the hydrodynamics of crystals.  The issue is puzzling because all the transport coefficients of crystals are known \cite{MPP72,FC76,GM83} and their Green-Kubo expressions have been deduced from statistical mechanics \cite{S97,MG20,MG21,H22,H23}.  Nevertheless, the precise relationship of these coefficients to the vacancy diffusion coefficient is still missing in the literature.  Another puzzling issue is that the coefficients that are associated with the transport of vacancies should vanish in the limit of a dilute-system of vacancies, although a diffusion coefficient is not expected to vanish even if the concentration of the diffusing species does.

In this paper, our purpose is to address these issues in the hydrodynamics of crystals with vacancies.  The key to resolve these issues and to let appear Fick's law for vacancies in the macroscopic equations of nonperfect monatomic crystals is to explicitly introduce the chemical potential of vacancies in the local-equilibrium approach to their macrodynamics.  The fact is that there exists a constraint relating the concentration of vacancies to the mass density and the strain tensor, so that the hydrodynamics of perfect crystals still applies but with macrofields modified by the chemical potential of vacancies.  We show that, as a consequence, the thermodynamic forces given by the gradients of the macrofields include a Fickian vacancy diffusion term, which does not vanish with the concentration of vacancies.  The microscopic foundations justifying the approach are provided.   In this way, the hydrodynamics of perfect crystals can be extended to apply to nonperfect crystals and to fully understand that the eighth mode is called the mode of vacancy diffusion.  The dispersion relations of the eight hydrodynamic modes are deduced from the coupled macroscopic equations to show that the eighth mode is indeed governed by a Fickian diffusion coefficient in nonperfect  monatomic crystals.  The results are confirmed by molecular dynamics simulations of the hard-sphere crystal with vacancies.  With these considerations, the vacancy diffusion coefficient is shown to be related to the previously known coefficients for the transport of vacancies.

The plan of the paper is the following.  The thermodynamics of crystals with vacancies is presented in Sec.~\ref{sec:thermo}.  The macroscopic time evolution of the crystal and the transport properties are introduced in Sec.~\ref{sec:time-evol}.  The microscopic foundations of the macroscopic description are given in Sec.~\ref{sec:micro}.  In Sec.~\ref{sec:hydro-eqs}, the macroscopic equations ruling the macrodynamics of nonperfect monatomic crystals are established and linearized around equilibrium.  The hydrodynamic modes, including the eighth mode of vacancy diffusion, are obtained in Sec.~\ref{sec:hydro-modes}.  Theory is applied to the hard-sphere crystal with vacancies in Sec.~\ref{sec:num_res_HS}, where the results of molecular dynamics simulations are reported.  Section~\ref{sec:conclusion} presents the conclusion and perspectives. Appendix~\ref{app:thermo} summarizes useful results about the equilibrium thermodynamics of a crystal with vacancies.  In App.~\ref{app:Gy}, we derive the Helfand moment of vacancies, which we use in the simulations.  Appendix~\ref{app:hydro-modes} gives the steps in the calculation of the hydrodynamic modes.  Appendix~\ref{app:S_vac} contains the definition of the vacancy spectral function and its hydrodynamic approximation.  Appendix~\ref{app:Markov} presents the Markovian jump model used in the case of the crystal with a single vacancy.

{\it Notations.} The spatial coordinates are labeled by the Latin indices $a, b, c, \ldots = x, y, z$, the hydrodynamic variables by the Greek indices $\alpha, \beta, \ldots$, and the particles or vacancies by the indices $i,j,\ldots = 1,2,\ldots$. Unless explicitly stated, Einstein's convention of summation over repeated indices is adopted. $k_{\rm B}$ denotes Boltzmann's constant and ${\rm i}=\sqrt{-1}$.


\section{Thermodynamics of crystals with vacancies}
\label{sec:thermo}

\subsection{Spatial distribution of vacancies}

Vacancies are defects in the crystal, where some lattice sites are unoccupied by the atom expected to occupy the site in the perfect crystal.  Crystals with vacancies can be simulated by molecular dynamics with periodic boundary conditions on the faces of some cubic or parallelepipedous domain of volume $V$.  The number $N$ of atoms is always conserved by this dynamics.  In the crystalline phase, the atoms form an ordered structure, having some number $N_0$ of lattice sites.  The crystal maintains its equilibrium lattice structure on average, so that the density of lattice sites is a constant quantity characteristic of the crystal and equal to its equilibrium value $n_0\equiv N_0/V=n_{{\rm eq},0}$.  In the presence of vacancies, the number $N$ of atoms is lower than the number $N_0$ of lattice sites.  The difference gives the number of vacancies, $N_{\rm v}\equiv N_0-N$.  

If the simulation proceeds out of equilibrium, the system is characterized by the density of atoms $n$, which varies in space and such that $N=\int_V n\, d{\bf r}$.  We may also introduce the mass density $\rho=m n$, where $m$ is the mass of each atom.  The equilibrium value of the mass density for the perfect crystal is equal to $\rho_{{\rm eq},0}=m n_{{\rm eq},0}$. 

In nonperfect crystals, vacancies are observed to jump between neighboring lattice sites, but their number $N_{\rm v}$ is conserved (if there is no creation of vacancy-interstitial pairs).  If the system is large enough, we may introduce a concentration or density of vacancies $n_{\rm v}$, which may vary in space under nonequilibrium conditions and such that $N_{\rm v}=\int_V n_{\rm v} \, d{\bf r}$.  (We note that the vacancy concentration or density is denoted $c=n_{\rm v}$ in Ref.~\cite{MG21}.) Equivalently, the spatial distribution of vacancies can be described by the local molar fraction of vacancies, 
\be
\label{y}
y \equiv \frac{n_{\rm v}}{n_{{\rm eq},0}} \, ,
\ee
which will play a central role in this paper.

Under strain, the crystal undergoes some displacement $u^a({\bf r},t)$, which may be a shear, a compression, or a dilatation.  In the linear regime of elasticity, the corresponding strain tensor is defined as
\be
\label{strain}
u^{ab} \equiv \frac{1}{2} \left( \nabla^a u^b + \nabla^b  u^a\right) .
\ee
Under nonequilibrium conditions, the local mass density of a crystal will differ from its mean equilibrium value $\rho_{{\rm eq},0}$ by two effects: The first is due to the presence of vacancies and the second to lattice strain, so that the local mass density $\rho$, the molar fraction of vacancies $y$, and the trace of the strain tensor $u^{aa}$ (i.e., the divergence of the  displacement field) are always related to each other by the constraint
\be
\label{constraint}
\rho = \rho_{{\rm eq},0} \left( 1 - y - u^{aa} \right).
\ee
Accordingly, the knowledge of two of these quantities determines the third and, in particular, the molar fraction of vacancies is given by
\be
\label{y-rho-u}
y = 1 - u^{aa} - \frac{\rho}{\rho_{{\rm eq},0}} \, ,
\ee
implying that any variation of the molar fraction of vacancies is related to variations in the trace of the strain tensor and the mass density as
\be
\label{dy-drho-du}
\delta y = - \delta u^{aa} - \frac{\delta\rho}{\rho_{{\rm eq},0}} \, .
\ee
We note that these relations can equivalently be expressed in terms of the density of atoms rather than the mass density with $n=n_{{\rm eq},0}(1-y-u^{aa})$, $y=1-u^{aa}-n/n_{{\rm eq},0}$, and $\delta y=-\delta u^{aa}-\delta n/n_{{\rm eq},0}$.  The relation $N=N_0-N_{\rm v}$ is always satisfied, because $N=\int_V n\, d{\bf r}=n_{{\rm eq},0}\left( V - \int_V y\, d{\bf r}- \int_V \boldsymbol{\nabla}\cdot{\bf u} \, d{\bf r}\right)=N_0-\int_V n_{\rm v}\, d{\bf r} - n_{{\rm eq},0}\oint_{\partial V} {\bf u}\cdot d\boldsymbol{\Sigma}=N_0-N_{\rm v}$, since $\oint_{\partial V} {\bf u}\cdot d\boldsymbol{\Sigma}=0$ by the periodic boundary conditions on the border $\partial V$ of the domain $V$.

Since the vacancies are conserved under the described conditions, we need to consider the local conservation equation of vacancies
\be
\label{y-eq}
\partial_t \, y + \nabla^a J_y^a = 0 \, ,
\ee
where $J_y^a$ is the current density of vacancies, in addition to the local conservation equations of mass, energy, momentum, and the evolution equation of the displacement field, which is the local order parameter of the crystalline phase.

\subsection{Local Gibbs and Euler thermodynamic relations}

For the purpose of formulating the thermodynamics of the crystal with vacancies, we have to introduce the chemical potential of vacancies $\mu_y$. In this respect, the Gibbs and Euler relations for the energy density $\epsilon$ read
\begin{align}
d\epsilon &= T \, ds + \mu \, d\rho + \mu_y \, dy + v^a \, d g^a + \phi^{ab} \, d u^{ab} \, , \label{Gibbs-e-y}\\
\epsilon &= T \, s + \mu \, \rho + \mu_y \, y + v^a \, g^a + \phi^{ab} \, u^{ab} - p \, , \label{Euler-e-y}
\end{align}
where $T$ is the temperature, $s$ is the entropy per unit volume (i.e., the entropy density), $\mu$ the chemical potential per unit mass, $v^a$ the velocity field, $g^a=\rho v^a$ the momentum density, $\phi^{ab}$ the excess of the reversible stress tensor $\sigma^{ab}=-p \,\delta^{ab}+\phi^{ab}$ with respect to the hydrostatic pressure~$p$.  The stress and excess stress tensors are symmetric by conservation of angular momentum, $\sigma^{ab}=\sigma^{ba}$ and $\phi^{ab}=\phi^{ba}$.  These thermodynamic relations hold in the frame of the laboratory, where the medium moves with the velocity ${\bf v}=(v^a)$.  In the frame moving with the element of the medium, the energy density and the chemical potential par unit mass take the values $\epsilon_0=\epsilon-\rho {\bf v}^2/2$ and $\mu_0=\mu+{\bf v}^2/2$, respectively.

The vacancy molar fraction can be eliminated by using the relation~(\ref{y-rho-u}) based on the constraint~(\ref{constraint}), so that the Gibbs and Euler relations can be expressed as
\begin{align}
d\epsilon &= T \, ds + \tilde\mu \, d\rho + v^a \, d g^a + \tilde\phi^{ab} \, d u^{ab} \, , \label{Gibbs-e}\\
\epsilon &= T \, s + \tilde\mu \, \rho + v^a \, g^a + \tilde\phi^{ab} \, u^{ab} - \tilde{p} \, , \label{Euler-e}
\end{align}
if the chemical potential per unit mass, the excess part of the stress tensor, and the pressure are redefined according to
\begin{align}
\tilde\mu &\equiv \mu -\frac{\mu_y}{\rho_{\rm eq,0}} \, , \label{mu-tilde}\\
\tilde\phi^{ab} &\equiv  \phi^{ab} - \mu_y \, \delta^{ab} \, , \label{phi-tilde}\\
\tilde{p} &\equiv p - \mu_y \, . \label{p-tilde}
\end{align}
We note that the redefined chemical potential~(\ref{mu-tilde}) can be written as $\tilde\mu=(\mu_{\rm a}-\mu_{\rm v})/m$ in terms of the chemical potential of the atoms $\mu_{\rm a}\equiv m\mu$, the chemical potential of the vacancies $\mu_{\rm v}\equiv\mu_y/n_{{\rm eq},0}$, and the mass $m$ of the atoms.  A remarkable property is that the reversible stress tensor remains unchanged through the redefinition of $\phi^{ab}$ and $p$,
\be\label{tilde-stress-tensor}
\tilde\sigma^{ab} = \sigma^{ab} \, .
\ee

The purpose of these redefinitions is that the molar fraction of vacancies has been eliminated from the thermodynamic relations, opening the way for a formulation where the thermodynamic force driving the flow of vacancies under nonequilibrium conditions does not explicitly appear, but  is taken into account by the thermodynamic forces of the other quantities.  The formulation is thus similar to the one considered in early works~\cite{MPP72,FC76}.  Nevertheless, the incorporation of the chemical potential of the vacancies into the redefined chemical potential, excess stress tensor, and pressure will have for consequence that the  Fickian term $-{\cal D}_y\nabla^a y$ of vacancy diffusion will now explicitly appear in the evolution equations of the hydrodynamics of the crystal, which was not the case in previous works.

At equilibrium, the thermodynamic quantities take uniform values, which are functions of the temperature, the atomic density, and the vacancy molar fraction.  The equilibrium thermodynamics in the limit of low vacancy density is summarized in App.~\ref{app:thermo}.


\section{Time evolution of the crystal}
\label{sec:time-evol}

\subsection{Local conservation equations}

The hydrodynamics of the crystal is ruled by local conservation equations for the mass, energy, momentum densities and the strain tensor, which can be expressed as
\be
\label{c-eq}
\partial_t \, c^{\alpha} + \nabla^a J_{c^{\alpha}}^a = 0
\ee
with the corresponding densities and current densities,
\be
(c^{\alpha}) = (\rho,\epsilon, g^b, u^{bc})
\quad\mbox{and}\quad
(J_{c^{\alpha}}^a) = (J_{\rho}^a,J_{\epsilon}^a, J_{g^b}^a, J_{u^{bc}}^a) \, .
\ee
The equation for the strain tensor can be deduced from the evolution equation for the displacement field, $\partial_t u^a+J_{u^a}=0$, which implies that the corresponding current density is given by $J_{u^{bc}}^a=\frac{1}{2}\left(\delta^{ab}\delta^{cd}+\delta^{ac}\delta^{bd}\right) J_{u^d}$.

In general, the current densities can be separated as $J_{c^{\alpha}}^a=\bar{J}_{c^{\alpha}}^a+{\cal J}_{c^{\alpha}}^a$ into a dissipativeless part $\bar{J}_{c^{\alpha}}^a$ and a dissipative part ${\cal J}_{c^{\alpha}}^a$.  The dissipativeless parts have known expressions~\cite{MG20,MG21}.  The current density of mass is equal to the momentum density, $J_{\rho}^a = g^a$.  Since the velocity field is defined as the momentum density divided by the mass density as $v^a\equiv g^a/\rho$, the local conservation equation of mass becomes the continuity equation, $\partial_t\, \rho + \nabla^a (\rho v^a)=0$.  In this respect, the dissipative part of the mass current density is equal to zero ${\cal J}_{\rho}^a =0$ and, thus, $J_{\rho}^a =\bar{J}_{\rho}^a = \rho v^a$.  For the momentum and energy current densities, their dissipativeless parts are given by $\bar{J}_{g^b}^a= \rho v^a v^b-\sigma^{ab}$ and $\bar{J}_{\epsilon}^a= \epsilon v^a -\sigma^{ab} v^b$, respectively, and their dissipative parts should be expressed in terms of the transport properties, as explained here below.  Moreover, the evolution equation for the displacement field can be expressed as $\partial_t u^a+\bar{J}_{u^a}+{\cal J}_{u^a}=0$ in terms of its dissipativeless part $\bar{J}_{u^a}=-v^a(1-u^{bb})$ and a dissipative part to be determined below.  We note that the vacancy current density introduced in Eq.~(\ref{y-eq}) has a similar decomposition $J_y^a=\bar{J}_y^a+{\cal J}_y^a$.  On the one hand, its dissipativeless part has the form $\bar{J}_y^a=y v^a$, so that $\bar{J}_y^a+\bar{J}_{u^a}=-v^a(1-y-u^{bb})=-\rho v^a/\rho_{{\rm eq},0}$ by Eq.~(\ref{constraint}).  On the other hand, its dissipative part should satisfy ${\cal J}_y^a= - {\cal J}_u^a$ in order to recover the local conservation equation of mass from the vacancy local conservation equation~(\ref{y-eq}) combined with the evolution equation for the  displacement field.

The dissipativeless parts of the current densities preserve entropy, i.e., they rule adiabatic (reversible) processes.  In contrast, their dissipative parts contribute to the entropy production rate.

\subsection{Entropy production and transport properties}

In order to obtain the dissipative parts of the current densities and the transport coefficients, we should identify the different contributions to the entropy production.  For this purpose, we first deduce the Gibbs relation for the entropy density from Eq.~(\ref{Gibbs-e}), giving
\be
k_{\rm B}^{-1} \, ds= \beta\, d\epsilon - \beta\tilde\mu \, d\rho - \beta v^b \, dg^b - \beta\tilde\phi^{bc} \, du^{bc} \, ,
\ee
where $\beta=(k_{\rm B}T)^{-1}$ is the inverse temperature.  By standard calculations of nonequilibrium thermodynamics \cite{GM84}, we can obtain the following time derivative for the entropy $S=\int_V s \, d{\bf r}$,
\begin{align}
\frac{1}{k_{\rm B}} \, \frac{dS}{dt} =& \int_V \Big[ \nabla^a\beta \, {\cal J}_{\epsilon}^a - \nabla^a(\beta\tilde\mu) \, {\cal J}_{\rho}^a - \nabla^a(\beta v^b) \, {\cal J}_{g^b}^a \nonumber\\
& - \nabla^a(\beta\tilde\phi^{bc}) \, {\cal J}_{u^{bc}}^a \Big] d{\bf r} \, .
\end{align}
As aforementioned, we have that ${\cal J}_{\rho}^a=0$.  Moreover, the gradients of products with $\beta$ can be expanded and all the  terms with the gradient of the inverse temperature can be gathered together, which leads to
\be
\frac{1}{k_{\rm B}} \, \frac{dS}{dt} = \int_V \Big[ \nabla^a\beta \, \tilde{\cal J}_{q}^a - (\beta\,\nabla^a v^b) \, {\cal J}_{g^b}^a - (\beta\,\nabla^a\tilde\phi^{ab}) \, {\cal J}_{u^b} \Big] d{\bf r}
\ee
with the heat current density
\be
\label{J-heat}
\tilde{\cal J}_{q}^a \equiv {\cal J}_{\epsilon}^a - v^b \, {\cal J}_{g^b}^a - \tilde\phi^{ab} \, {\cal J}_{u^b} \, ,
\ee
obtained by  again using the relation ${\cal J}_{u^{bc}}^a=\frac{1}{2}\left(\delta^{ab}\delta^{cd}+\delta^{ac}\delta^{bd}\right) {\cal J}_{u^d}$.  Since the system is isolated, the time derivative of the entropy gives the rate of entropy production  caused by the dissipative current densities.

The entropy production rate can be equivalently expressed as
\be
\frac{dS}{dt} = \int_V \Big( {\cal A}_{q}^a \, \tilde{\cal J}_{q}^a + {\cal A}_{g^b}^a \, {\cal J}_{g^b}^a+ \tilde{\cal A}_{u^b} \, {\cal J}_{u^b} \Big) d{\bf r}
\ee
in terms of the thermodynamic forces or affinities, which are here identified as
\begin{align}
{\cal A}_{q}^a &\equiv k_{\rm B} \nabla^a \beta \, , \label{A-q} \\
{\cal A}_{g^b}^a &\equiv - k_{\rm B}\, \beta \, \nabla^a v^b \, , \label{A-g} \\
\tilde{\cal A}_{u^b} &\equiv - k_{\rm B}\, \beta \, \nabla^a \tilde\phi^{ab} \, . \label{tilde-A-u}
\end{align}
Because of Eq.~(\ref{phi-tilde}), the thermodynamic force associated with the displacement includes a mechanical contribution due to the excess of stress and another contribution due to the vacancies:
\be
\tilde{\cal A}_{u^b} = {\cal A}_{u^b} + k_{\rm B}\, \beta \, \nabla^b \mu_y
\quad\mbox{with}\quad
{\cal A}_{u^b} \equiv - k_{\rm B}\, \beta \, \nabla^a \phi^{ab} \, .
\label{A-u}
\ee
The affinities~(\ref{A-q}) and~(\ref{A-g}) driving heat and momentum transports do not have a tilde because they are the same as in fluids and perfect crystals.

Now, if the macrofields have small enough gradients, the dissipative parts of the current densities are given by linear combinations of the affinities.  The coefficients of these linear combinations define the possible transport properties of the crystal.  In centrosymmetric crystals, where rank-three tensors are equal to zero~\cite{MG21}, we find the following linear combinations,
\begin{align}
\tilde{\cal J}_{q}^a &= \tilde{\cal L}_{qq}^{ab} \, {\cal A}_{q}^b +  \tilde{\cal L}_{qu}^{ab} \, \tilde{\cal A}_{u^b} \, , \label{J-q-L-A} \\
{\cal J}_{u^a} &= \tilde{\cal L}_{uq}^{ab} \, {\cal A}_{q}^b + {\cal L}_{uu}^{ab} \, \tilde{\cal A}_{u^b} \, , \label{J-u-L-A} \\
{\cal J}_{g^b}^a &= {\cal L}_{gg}^{abcd} \, {\cal A}_{g^d}^c \, , \label{J-g-L-A}
\end{align}
in terms of linear response coefficients ${\cal L}^{ab}_{c^\alpha c^\beta}$, which are proportional to the transport coefficients.
The tildes denote the coefficients and the quantities that are modified by the vacancy chemical potential $\mu_y$.
The rank-four tensor of viscosity coefficients is given by $\eta^{abcd} \equiv {\cal L}_{gg}^{abcd}/T$, so that we recover the expected phenomenological relation between the dissipative part of the momentum current density and the gradient of velocity, ${\cal J}_{g^b}^a = -\eta^{abcd} \nabla^c v^d$.  The interpretation of the rank-two tensors requires to obtain their microscopic expression in terms of Green-Kubo formulas~\cite{G52,G54,K57}, as explained in the following section.


\section{Microscopic foundations}
\label{sec:micro}

\subsection{Hamiltonian dynamics and microscopic quantities}

The motion of atoms composing the crystal is ruled by Hamiltonian dynamics with periodic boundary conditions on the domain of volume $V$ considered.  The densities and current densities of the locally conserved quantities can be expressed in terms of the positions and momenta of the atoms \cite{MG20,MG21}.  These microscopic observables $\hat{c}^{\alpha}$ and $\hat{J}_{c^{\alpha}}^a$ obey local conservation equations similar to Eq.~(\ref{c-eq}).  The microscopic forms of the displacement field $\hat{u}^a$ and the vacancy concentration or density $\hat{c}=\hat{n}_{\rm v}$ are also known \cite{S97,MG21}.  In particular, in the molecular dynamics of the crystal, the positions  ${\bf r}_{{\rm v},i}(t)$ of the vacancies can be determined at any time $t$ by the positions of the lattice sites that are not occupied by an atom, allowing us to define the microscopic molar fraction of vacancies as
\begin{align}
\label{y-micro}
\hat{y}({\bf r},t) \equiv n_{{\rm eq},0}^{-1}\,\hat{c}({\bf r},t) = n_{{\rm eq},0}^{-1}\sum_{i=1}^{N_{\rm v}} \delta[{\bf r}-{\bf r}_{{\rm v},i}(t)] \, .
\end{align}
We note that an alternative definition would consist in replacing the Dirac delta distribution in Eq.~(\ref{y-micro}) by a smooth function with a peak having a width of the order of the size of a lattice site \cite{MG21}.  Nevertheless, we shall use the definition~(\ref{y-micro}), in particular, for the simplicity it provides in computing the quantities of interest. Furthermore, the constraint~(\ref{constraint}) and the relation~(\ref{y-rho-u}) are assumed to hold between the microscopic quantities $\hat{\rho}$, $\hat{u}^{aa}=\nabla^a\hat{u}^a$, and~$\hat{y}$.

\subsection{Local-equilibrium probability distribution}

In the local-equilibrium approach \cite{M58,McL63,R66,P68,OL79,S14}, the phase-space probability distribution is first surmised by a Gibbsian distribution expressed in terms of fields, which vary in space and time, and called the local-equilibrium probability distribution.  In a second step, this surmised probability distribution is related to the exact probability distribution ruled by Liouville's equation in phase space.

For a crystal with vacancies, the local-equilibrium probability distribution can be defined {\it a priori} as
\begin{align}
\label{p-leq}
{\cal P}_{\rm leq}(\Gamma;\boldsymbol{\lambda}) =&\ \frac{1}{\Delta\Gamma} \, \exp \Big[ -\Omega - \int_V \big( \lambda_\epsilon \,\hat{\epsilon} + \lambda_\rho\, \hat{\rho} + \lambda_y\, \hat{y} \nonumber\\
&+\; \lambda_{g^a}\, \hat{g}^a +  \lambda_{u^{ab}}\, \hat{u}^{ab} \big) \, d{\bf r} \Big] \, ,
\end{align}
where $\Gamma\in{\mathbb R}^{6N}$ are the phase-space coordinates of positions and momenta of the $N$ atoms in the system, $\Delta\Gamma$ is the quantal phase-space volume, $\Omega$ a normalization constant, and the quantities $\lambda^{\alpha}$ are fields conjugate to the microscopic densities.  In this definition, the vacancy molar fraction $\hat{y}$ is considered as an independent variable.  At leading order in an expansion in the gradients of the macrofields, the conjugate fields are given by
\begin{align}\label{conjugated_fields}
\lambda_{\epsilon} &= \beta + O(\nabla^2) \, ,& 
\lambda_{\rho} &= -\beta\, \mu + O(\nabla^2) \, , \nonumber\\
\lambda_{y} &= -\beta\, \mu_y + O(\nabla^2) \, , &
\lambda_{g^a} &= -\beta\, v^a+ O(\nabla^2) \, , \nonumber\\ 
\lambda_{u^{ab}} &= -\beta\, \phi^{ab}+ O(\nabla^2) \, ,
\end{align}
and $\Omega=\int_V \beta \, p \, d{\bf r}+O(\nabla^2)$, as expressed in terms of the thermodynamic quantities appearing in the Gibbs and Euler relations~(\ref{Gibbs-e-y}) and~(\ref{Euler-e-y}).  Since the constraint~(\ref{constraint}) holds at the microscopic level of description as well as at the macroscale, the vacancy molar fraction $\hat{y}$ can be eliminated using Eq.~(\ref{y-rho-u}) 
and the local-equilibrium probability distribution may be expressed as
\begin{align}\label{p-leq-new}
{\cal P}_{\rm leq}(\Gamma;\tilde{\boldsymbol{\lambda}}) =&\ \frac{1}{\Delta\Gamma} \, \exp \Big[ -\tilde\Omega - \int_V \big( \lambda_\epsilon \,\hat{\epsilon} + \tilde\lambda_\rho\, \hat{\rho}  \nonumber\\
&+ \; \lambda_{g^a}\, \hat{g}^a +  \tilde\lambda_{u^{ab}}\, \hat{u}^{ab} \big) \, d{\bf r} \Big] 
\end{align}
with
\begin{align}
\tilde\lambda_{\rho} &\equiv \lambda_{\rho} -\frac{\lambda_y}{\rho_{{\rm eq},0}} = - \beta\, \tilde\mu + O(\nabla^2) \, , \label{tilde-lambda-rho} \\
\tilde\lambda_{u^{ab}} &\equiv \lambda_{u^{ab}} - \lambda_y \, \delta^{ab} = -\beta \, \tilde\phi^{ab} + O(\nabla^2) \, , \label{tilde-lambda-u} \\
\tilde\Omega &\equiv \Omega + \int_V \lambda_y \, d{\bf r} = \int_V \beta \, \tilde{p} \, d{\bf r} + O(\nabla^2)  \, , \label{tilde-Omega}
\end{align}
which are analogous to Eqs.~(\ref{mu-tilde})-(\ref{p-tilde}) at the macroscale.

In the local-equilibrium approach, the macroscopic densities are given by the statistical averages of the microscopic densities over the local-equilibrium probability distribution~(\ref{p-leq-new}), $c^{\alpha}=\langle\hat{c}^{\alpha}\rangle_{{\rm leq},\tilde{\boldsymbol{\lambda}}}$. The entropy is obtained as the Legendre transform $S({\bf c})= k_{\rm B} \, {\rm inf}_{\tilde{\boldsymbol{\lambda}}}\left[{\tilde\Omega}(\tilde{\boldsymbol{\lambda}})+\int_V \tilde\lambda^{\alpha}c^{\alpha}\, d{\bf r}\right]$.  In this way, the Gibbs and Euler relations~(\ref{Gibbs-e}) and~(\ref{Euler-e}) can be deduced from the microscopic level of description.

After taking into account the constraint~(\ref{constraint}), the local-equilibrium probability distribution has now the same form as the one used in Refs.~\cite{MG20,MG21,MG23} to obtain the macrodynamics of crystals and other matter phases with broken continuous symmetries.  Therefore, the methods developed in these references can here be applied to determine the transport properties of crystals with vacancies.

\subsection{Microscopically exact probability distribution}

The  crucial point of the local-equilibrium approach is that the exact probability distribution ${\cal P}_t(\Gamma)$ obeying Liouville's equation in phase space is related to the local-equilibrium probability distribution~(\ref{p-leq-new}) by
\be
\label{p-exact}
{\cal P}_t(\Gamma) = {\cal P}_{\rm leq}(\Gamma;\tilde{\boldsymbol{\lambda}}_t) \, {\rm e}^{\Sigma_t(\Gamma)}
\ee
with the following quantity,
\be
\label{Sigma}
\Sigma_t(\Gamma) \equiv \int_0^t d\tau \, \partial_{\tau}\Big[ {\tilde\Omega}(\tilde{\boldsymbol{\lambda}}_{\tau}) + \int_V \tilde{\lambda}_{\tau}^{\alpha} \, \hat{c}^{\alpha}_{\tau-t} \, d{\bf r} \Big] \, ,
\ee
where $\tilde{\boldsymbol{\lambda}}_{\tau}=(\tilde{\lambda}_{\tau}^{\alpha})$ denotes the modified conjugate fields at time $\tau$ and $\hat{c}^{\alpha}_{\tau-t}$ the microscopic densities for the particles at time ${\tau-t}$ with respect to initial conditions fixed at the phase-space point $\Gamma$ \cite{McL63,S14,MG20}.  The quantity~(\ref{Sigma}) plays a key role to obtain the dissipative parts of the current densities and the entropy production rate.  According to the relation~(\ref{p-exact}), the mean value of an observable $A(\Gamma)$ with respect to the exact probability distribution can be expressed as
\be
\label{av-A-exact-leq}
\langle A(\Gamma) \rangle_t  = \langle A(\Gamma)\, {\rm e}^{\Sigma_t(\Gamma)}\rangle_{{\rm leq},{\tilde{\boldsymbol{\lambda}}}_t}
\ee
in terms of the local-equilibrium probability distribution~(\ref{p-leq-new}).  In particular, the integral fluctuation relation $\langle{\rm e}^{-\Sigma_t(\Gamma)}\rangle_t=1$ holds, which implies the non-negativity of the entropy production by Jensen's inequality, $S({\bf c}_t)-S({\bf c}_0)=k_{\rm B}\, \langle\Sigma_t(\Gamma)\rangle_t \geq 0$ \cite{S14}.

\subsection{Dissipativeless and dissipative density currents}

Taking $A(\Gamma)=\hat{J}_{c^\alpha}^a({\bf r};\Gamma)$ in Eq.~(\ref{av-A-exact-leq}), the mean values of the current densities can be expressed in terms of statistical averages over the local-equilibrium distribution.   We can identify the dissipativeless and dissipative mean current densities as
\begin{align}
\bar{J}^{a}_{c^{\alpha}}(\mathbf{r},t) &\equiv \langle \hat{J}^a_{c^{\alpha}}(\mathbf{r};\Gamma)\rangle_{{\rm leq},\tilde{\boldsymbol{\lambda}}_t} \, , \label{dfn-J-bar}\\
{\mathcal{J}}^a_{c^{\alpha}}(\mathbf{r},t) &\equiv \langle \hat{J}^a_{c^{\alpha}}(\mathbf{r};\Gamma)\big[{\rm e}^{\Sigma_t(\Gamma)}-1\big]\rangle_{{\rm leq},\tilde{\boldsymbol{\lambda}}_t}  \nonumber\\
&= \langle \hat{J}^{a}_{{c}^\alpha}(\mathbf{r};\Gamma)\,\Sigma_t(\Gamma)\rangle_{{\rm leq},\tilde{\boldsymbol{\lambda}}_t} + O(\Sigma_t^2) \, ,
\label{dfn-J-cal}
\end{align}
respectively, because the dissipativeless parts~(\ref{dfn-J-bar}) do not contribute to the time derivative of the entropy~\cite{OL79,S14,MG20}.

The method  to obtain the dissipative current densities consists in evaluating the quantity~(\ref{Sigma}) at leading order in the deviations with respect to local equilibrium.  The calculation is similar to the one performed in Refs.~\cite{MG20,MG21,MG23}, but here using the energy current density $\delta\hat{\tilde{J}}_\epsilon^a = \delta\hat{J}_\epsilon^a + \mu_y \delta\hat{J}_{u^a}$, which is modified to include the vacancy chemical potential~$\mu_y$.  Another modification is that the calculation here uses the reversible stress tensor~(\ref{tilde-stress-tensor}) after elimination of the vacancy molar fraction $y$ to get $\tilde\sigma^{ab}(\epsilon_0,\rho,u^{cd})=\sigma^{ab}[\epsilon_0,\rho,u^{cd},y(\rho,u^{ee})]$ with $y(\rho,u^{ee})$ given by Eq.~(\ref{y-rho-u}).  After the calculation, we find the following expression,
\begin{align}
\Sigma_t(\Gamma) = & \int_0^t d\tau \int_V d{\bf r}\, \Big[{\nabla}^a\beta_\tau\, \delta\hat{\tilde{J}}^{\prime a}_{\epsilon,{\tau-t}} -(\beta_\tau{\nabla}^a{v}^b_\tau)\, \delta\hat{J}^{\prime a}_{g^b,\tau-t} \nonumber\\
& - \; (\beta_\tau{\nabla}^a\tilde{\phi}^{ab}_\tau)\,\delta\hat{J}^{\prime}_{u^b,\tau-t}\Big] ,
\end{align}
where
\begin{align}
\delta\hat{\tilde{J}}^{\prime a}_{\epsilon} &\equiv \delta\hat{J}^{a}_{\epsilon}-\rho^{-1} (\epsilon_0+p) \, \delta\hat{g}^a + \mu_y \left(\delta\hat{J}_{u^a}+ \rho^{-1} \, \delta\hat{g}^a\right) , \label{delta_j_e} \\
\delta\hat{J}^{\prime a}_{g^b} &\equiv \delta\hat{J}^{ a}_{g^b} +R^{ab}_{\epsilon}\, \delta\hat{\epsilon} +R^{ab}_{\rho}\,  \delta\hat{\rho} + R^{abcd}_{u} \, \delta\hat{u}^{cd}\, , \label{delta_j_g} \\
\delta\hat{J}^{\prime}_{u^b} &\equiv \delta\hat{J}_{u^b}+ \rho^{-1} \, \delta\hat{g}^b \, , \label{delta_j_u}
\end{align}
with
\begin{align}
R^{ab}_{\epsilon} &\equiv \left(\frac{\partial\sigma^{ab}}{\partial\epsilon_0}\right)_{\rho,y,\boldsymbol{\mathsf u}} \, , \\
R^{ab}_{\rho} &\equiv \left(\frac{\partial\sigma^{ab}}{\partial\rho}\right)_{\epsilon_0,y,\boldsymbol{\mathsf u}} -\rho^{-1}\left(\frac{\partial\sigma^{ab}}{\partial y}\right)_{\epsilon_0,\rho,\boldsymbol{\mathsf u}} \, , \\
R^{abcd}_{u} &\equiv \left(\frac{\partial\sigma^{ab}}{\partial u^{cd}}\right)_{\epsilon_0,\rho,y,\boldsymbol{\mathsf u^\prime}} - \left(\frac{\partial\sigma^{ab}}{\partial y}\right)_{\epsilon_0,\rho,\boldsymbol{\mathsf u}} \delta^{cd} \, , 
\end{align}
$\boldsymbol{\mathsf u}$ denoting all the elements of the tensor $(u^{cd})$, $\boldsymbol{\mathsf u^\prime}$ the elements other than the one involved in the partial derivative, and $\rho=\rho_{{\rm eq},0}$.

\subsection{Green-Kubo formulas for transport coefficients}

The microscopic global currents are defined by
\be
\delta{\mathbb J}_{c^\alpha}^a(t) \equiv \int_V \delta\hat{J}_{c^\alpha}^a({\bf r},t) \, d{\bf r} \, .
\ee
If the computations are carried out using the microcanonical ensemble, where the total momentum, energy, and mass are conserved, the terms involving these conserved quantities in the microscopic global currents are equal to zero, i.e., $\delta{P}^a = \int_V \delta\hat{g}^a \, d{\bf r}=0$, $\delta{E} = \int_V \delta\hat{\epsilon} \, d{\bf r}=0$, and $\delta{M} = \int_V \delta\hat{\rho} \, d{\bf r}=0$.  In addition, we have that $\delta{U}^{cd}= \int_V \delta\hat{u}^{cd} \, d{\bf r}=0$ in this ensemble, as shown in Ref.~\cite{MG24a}.  Hence, the microscopic global currents to consider are
\be
\label{tilde-J_e}
\delta\tilde{\mathbb J}^{a}_{\epsilon} = \delta{\mathbb J}^{a}_{\epsilon} + \mu_y \,\delta{\mathbb J}_{u^a} \, , 
\ee
$\delta{\mathbb J}^{a}_{g^b}$, and $\delta{\mathbb J}_{u^b}$.
We note that the global current of energy is modified due to the vacancy chemical potential $\mu_y$.
Furthermore, the microscopic global current of vacancies is given by
\be
\label{J_y-J_u}
\delta{\mathbb J}_{y}^a = - \delta{\mathbb J}_{u^a}
\ee
because $\delta\hat{J}_{y}^a=-\delta\hat{J}_{u^a}$.

Therefore, the linear response coefficients are given by the following Green-Kubo formulas, 
\begin{align}
\tilde{\cal L}_{qq}^{ab} &= \lim_{V\rightarrow \infty} \frac{1}{k_{\rm B} V}\int_0^{\infty}{\rm d}t \, \langle \delta \tilde{\mathbb J}^{a}_{\epsilon}(t)\, \delta \tilde{\mathbb J}^{b}_{\epsilon}(0)\rangle_{\rm eq} \, ,\label{tilde-kappa}\\
\tilde{\cal L}_{qu}^{ab} &= \lim_{V\rightarrow \infty}  \frac{1}{k_{\rm B} V}\int_0^{\infty}{\rm d}t \, \langle \delta \tilde{\mathbb J}^{a}_{\epsilon}(t)\,  \delta {\mathbb J}_{u^b}(0)\rangle_{\rm eq} \, , \label{tilde-xi}\\
{\cal L}_{uu}^{ab} &= \lim_{V\rightarrow \infty}  \frac{1}{k_{\rm B} V}\int_0^{\infty}{\rm d}t \, \langle \delta {\mathbb J}_{u^a}(t) \, \delta {\mathbb J}_{u^b}(0)\rangle_{\rm eq} \, , \label{zeta}\\
{\cal L}_{gg}^{abcd} &= \lim_{V\rightarrow \infty}  \frac{1}{k_{\rm B} V} \int_0^{\infty}{\rm d}t \, \langle \delta {\mathbb J}^{a}_{g^b}(t)\, \delta  {\mathbb J}^{c}_{g^d}(0)\rangle_{\rm eq} \, . \label{eta} \qquad
\end{align}
We can also introduce the coefficients ${\cal L}_{qq}^{ab}$ and ${\cal L}_{qu}^{ab}$ defined by Green-Kubo formulas similar to Eqs.~(\ref{tilde-kappa}) and~(\ref{tilde-xi}) but with $\delta\tilde{\mathbb J}^{a}_{\epsilon}$ replaced by $\delta{\mathbb J}^{a}_{\epsilon}$. Since the microscopic dynamics and the global currents are here symmetric under time reversal, the following Onsager reciprocal relations are satisfied,
\be
\label{Onsager}
{\cal L}_{qq}^{ab}= {\cal L}_{qq}^{ba} \, , \ \ \;
{\cal L}_{qu}^{ab} = {\cal L}_{uq}^{ba} \, , \ \ \;
{\cal L}_{uu}^{ab} = {\cal L}_{uu}^{ba} \, , \ \ \;
{\cal L}_{gg}^{abcd} = {\cal L}_{gg}^{cdab} .
\ee
Here, the phenomenological transport coefficients are defined as
\begin{align}
 \kappa^{ab} &\equiv T^{-2} \, {\cal L}_{qq}^{ab} \, , &
\xi^{ab} &\equiv T^{-1} \, {\cal L}_{qu}^{ab} \, , \nonumber\\
\zeta^{ab} &\equiv T^{-1} \, {\cal L}_{uu}^{ab} \, , &
\eta^{abcd} &\equiv T^{-1} \, {\cal L}_{gg}^{abcd} \, ,
\label{coeff-L}
\end{align}
$\kappa^{ab}$ being the heat conductivities~\cite{MG24a}, $\xi^{ab}$ the coefficients coupling heat and vacancy transports, $\zeta^{ab}$ those ruling vacancy transport, and $\eta^{abcd}$ the viscosities.

Because of Eq.~(\ref{tilde-J_e}), the coefficients~(\ref{tilde-kappa}) and~(\ref{tilde-xi}) can be expressed as follows,
\begin{align}
\tilde{\cal L}_{qq}^{ab} &= {\cal L}_{qq}^{ab} + \mu_y \, {\cal L}_{qu}^{ab} + \mu_y \, {\cal L}_{uq}^{ab} + \mu_y^2 \, {\cal L}_{uu}^{ab} \nonumber\\
&= T^2\left(\kappa^{ab} + \frac{\mu_y}{T} \, \xi^{ab} +  \frac{\mu_y}{T} \, \xi^{ba} + \frac{\mu_y^2}{T} \, \zeta^{ab}\right) , \label{tilde-L_qq}\\
\tilde{\cal L}_{qu}^{ab} &= {\cal L}_{qu}^{ab} + \mu_y \, {\cal L}_{uu}^{ab} 
= T\left(\xi^{ab} + \mu_y \, \zeta^{ab}\right) , \label{tilde-L_qu}
\end{align}
in terms of the phenomenological transport coefficients.

\subsection{Einstein-Helfand formulas for transport coefficients}

Furthermore, we may introduce the Helfand moments
\be
\label{Helfand}
{\mathbb G}_{c^\alpha}^a(t) \equiv \int_0^t \delta{\mathbb J}_{c^\alpha}^a(\tau) \, d\tau \, .
\ee
Because the   identity
\be
\int_0^{\infty} dt \, \langle \delta {\mathbb J}^{a}_{c^\alpha}(t)\, \delta {\mathbb J}^{b}_{c^\beta}(0)\rangle_{\text{eq}} = \lim_{t\to\infty} \frac{1}{2t} \, \langle {\mathbb G}^{a}_{c^\alpha}(t)\, {\mathbb G}^{b}_{c^\beta}(t)\rangle_{\text{eq}}
\ee
holds if the two global currents have the same parity under time reversal, the Green-Kubo formulas can be transformed into Einstein-Helfand formulas~\cite{E26,H60}, giving the transport coefficients as follows,
\be
{\cal L}_{c^\alpha c^\beta}^{ab}  = \lim_{V\to\infty} \lim_{t\to\infty} \frac{1}{2t k_{\rm B}V} \, \langle {\mathbb G}^{a}_{c^\alpha}(t)\, {\mathbb G}^{b}_{c^\beta}(t)\rangle_{\text{eq}} \, .
\ee
Moreover,  according to Eq.~(\ref{J_y-J_u}), the Helfand moment for the vacancies  should satisfy
\be
{\mathbb G}_{y}^a(t) = - {\mathbb G}_{u^a}(t) \, .
\ee
Therefore, we have that
\begin{align}
\kappa^{ab} &= \lim_{V\to\infty} \lim_{t\to\infty} \frac{1}{2t k_{\rm B} T^2 V} \, \langle {\mathbb G}^{a}_{\epsilon}(t)\,  {\mathbb G}_{\epsilon}^b(t)\rangle_{\rm eq} \, , \label{kappa-H}\\
\xi^{ab} &= - \lim_{V\to\infty} \lim_{t\to\infty} \frac{1}{2t k_{\rm B} T V} \, \langle {\mathbb G}^{a}_{\epsilon}(t)\,  {\mathbb G}_{y}^b(t)\rangle_{\rm eq} \, , \label{xi-H-y}\\
\zeta^{ab} &= \lim_{V\to\infty} \lim_{t\to\infty} \frac{1}{2t k_{\rm B} T V} \, \langle {\mathbb G}_{y}^a(t) \, {\mathbb G}_{y}^b(t)\rangle_{\rm eq} \, . \label{zeta-H-y}
\end{align}
As shown in App.~\ref{app:Gy}, the Helfand moment of the vacancies can be expressed in terms of their position in the lattice. We note that Einstein-Helfand formulas also exist for the viscosities \cite{MG24a}.


\section{The equations of crystal hydrodynamics}
\label{sec:hydro-eqs}

\subsection{The equations with their dissipative current densities}

Gathering all the previous results, the macroscopic equations of the dissipative hydrodynamics of monatomic crystals are given by
\begin{align}
\partial_t \rho  +\nabla^a(\rho v^a) &= 0 \, , \label{hydro-eq-rho}\\
\partial_t y  +\nabla^a(y \, v^a + {\mathcal J}^a_y) &= 0 \, , \label{hydro-eq-y}\\
\partial_t \epsilon+\nabla^a( \epsilon\, v^a - \sigma^{ab} v^b + {\mathcal J}_{\epsilon}^a) &= 0 \, , \label{hydro-eq-eps}\\
\partial_t (\rho v^b) +\nabla^a( \rho v^a v^b -\sigma^{ab}  + {\mathcal J}^a_{g^b}) &= 0 \, , \label{hydro-eq-g}\\
\partial_t u^a -v^a (1-u^{bb}) + {\mathcal J}_{u^a}  &= 0 \, , \label{hydro-eq-u}
\end{align}
where the dissipative current densities take the following forms,
\begin{align}
{\cal J}_{\epsilon}^a =& -\left(\kappa^{ab}+ \frac{\mu_y}{T} \, \xi^{ab}\right) \nabla^b T - \xi^{ab} \, \nabla^c\phi^{cb} + \xi^{ab} \, \nabla^b \mu_y \nonumber\\
& + \;  v^b \, {\cal J}_{g^b}^a + \phi^{ab} \, {\cal J}_{u^b} \, , \label{JD-e-fin}\\
{\cal J}_{u^a} =& -\frac{1}{T} \left(\xi^{ba} + \mu_y \, \zeta^{ab}\right) \nabla^b T - \zeta^{ab} \, \nabla^c\phi^{cb} + \zeta^{ab} \, \nabla^b \mu_y \nonumber\\
=& - {\cal J}_{y}^a  \, , \label{JD-u-fin} \\
{\cal J}_{g^b}^a =& -\eta^{abcd} \, \nabla^c v^d \, , \label{JD-g-fin}
\end{align}
as a consequence of Eqs.~(\ref{J-q-L-A})-(\ref{J-g-L-A}) expressing the dissipative current densities in terms of the affinities~(\ref{A-q})-(\ref{A-u}) and the transport coefficients~(\ref{coeff-L}).  We note that Eq.~(\ref{JD-e-fin}) is obtained using Eq.~(\ref{J-heat}).  

The modifications of the hydrodynamic equations due to the vacancy chemical potential have been explicitly calculated.  In the presence of a gradient of vacancy chemical potential, there is an extra contribution in the dissipative current densities of energy and the vacancies.

In cubic crystals, the rank-two tensors are diagonal, so that
\be
\kappa^{ab} = \kappa \, \delta^{ab} \, , \quad
\xi^{ab} = \xi \, \delta^{ab} \, , \quad
\mbox{and}\quad
\zeta^{ab} = \zeta \, \delta^{ab} \, ,
\label{eq:cubsym}
\ee
which define the heat conductivity $\kappa$, the coefficient~$\xi$ of vacancy thermodiffusion, and the vacancy conductivity~$\zeta$.  Moreover, the rank-four tensor of viscosities can be expressed in terms of three coefficients, $\eta_{11}$, $\eta_{12}$, and $\eta_{44}$ in Voigt's notations.

The vacancy chemical potential $\mu_y$ depends on the temperature, the vacancy molar fraction, and the excess stress.  Therefore, we have the following expansion for the gradient of the vacancy chemical potential
\be
\nabla^a \mu_y = \left(\frac{\partial\mu_y}{\partial T}\right)_{y,\boldsymbol{\phi}} \nabla^a T + \left(\frac{\partial\mu_y}{\partial y}\right)_{T,\boldsymbol{\phi}} \nabla^a y + \mu_{y,\boldsymbol{\phi}} \, \nabla^a \phi^{bb}
\ee
with
\be
\mu_{y,\boldsymbol{\phi}} \equiv \frac{1}{3}\, {\rm tr}\left(\frac{\partial\mu_y}{\partial\boldsymbol{\phi}}\right)_{T,y} \, ,
\ee
$\boldsymbol{\phi}$ denoting the tensor $(\phi^{cd})$.
Substituting into Eqs.~(\ref{JD-e-fin}) and~(\ref{JD-u-fin}), we get
\begin{align}
{\cal J}_{\epsilon}^a =& -\kappa^{\prime} \, \nabla^a T - \xi \, \nabla^b\phi^{ba} + \xi \, \mu_{y,\boldsymbol{\phi}} \, \nabla^a\phi^{bb} + {\cal K}_y \, \nabla^a y  \nonumber\\
&+ v^b \, {\cal J}_{g^b}^a + \phi^{ab} \, {\cal J}_{u^b} \, , \label{JD-e-fin-cubic} \\
{\cal J}_{u^a} =& -\frac{\xi^{\prime}}{T} \, \nabla^a T - \zeta \, \nabla^b\phi^{ba} + \zeta \,\mu_{y,\boldsymbol{\phi}} \,  \nabla^a\phi^{bb} + {\cal D}_y \, \nabla^a y \nonumber\\
 =& - {\cal J}_{y}^a \, , \label{JD-u-fin-cubic}
\end{align}
where
\begin{align}
\kappa^{\prime} &\equiv \kappa - T \left[\frac{\partial(\mu_y/T)}{\partial T}\right]_{y,\boldsymbol{\phi}} \xi \, , \label{kappa-prime}\\
\xi^{\prime} &\equiv \xi - T^2 \left[\frac{\partial(\mu_y/T)}{\partial T}\right]_{y,\boldsymbol{\phi}} \zeta \, , \label{xi-prime}\\
{\cal K}_y &\equiv \left(\frac{\partial\mu_y}{\partial y}\right)_{T,\boldsymbol{\phi}} \xi \, , \label{K_y}\\
{\cal D}_y &\equiv \left(\frac{\partial\mu_y}{\partial y}\right)_{T,\boldsymbol{\phi}} \zeta \, . \label{D_y}
\end{align}

A remarkable result is that the local conservation equation of vacancies~(\ref{hydro-eq-y}) can now be written as
\be
\label{hydro-eq-y-2}
\partial_t \, y  +\nabla^a({\cal V}_y^a \, y  -{\cal D}_y \nabla^a y) = 0 \, ,
\ee
where $-{\cal D}_y\nabla^a y$ is the term expressing Fick's law for vacancy diffusion and
\be
{\cal V}_y^a \equiv v^a + \frac{\xi^{\prime}}{T\, y} \, \nabla^a T + \frac{\zeta}{y} \left( \nabla^b\phi^{ba} - \mu_{y,\boldsymbol{\phi}} \,  \nabla^a\phi^{bb}\right) \label{drift-velocity-y}
\ee
can be interpreted as the overall velocity driving the motion of the vacancies not only by the velocity field $v^a$ of the medium, but also by the temperature gradient $\nabla^a T$ and the gradient of the excess stress tensor $\nabla^b\phi^{cd}$, which can be generated by the mechanical deformation of the crystal.  In the limit of arbitrarily low vacancy density, we expect that individual vacancies are still subjected to these thermal and mechanical driving forces, whereupon we can infer that $\zeta$ and $\xi$ should be proportional to the vacancy molar fraction $y$ in the limit $y\to 0$.  Furthermore, Eq.~(\ref{hydro-eq-y-2}) now contains a genuine diffusion term and the coefficient ${\cal D}_y$ can thus be identified as the vacancy diffusion coefficient.

\subsection{The linearized equations}

The dissipative current densities (\ref{JD-e-fin-cubic})-(\ref{JD-u-fin-cubic}) are linearized around the equilibrium rest state.  In this regard, the last two terms drop in Eq.~(\ref{JD-e-fin-cubic}) for the dissipative current density of energy, because $v^b$ and $\phi^{ab}$ are already deviations with respect equilibrium and   each one of them multiply a nonequilibrium current density.  Around the equilibrium rest state, all the coefficients take their constant and uniform equilibrium value.

We may introduce the specific entropy~${\mathfrak s}\equiv s/\rho$ (i.e., the entropy per unit mass) such that $S=\int_V \rho\, {\mathfrak s}\, d{\bf r}$ and having  deviations given by the following Gibbs relation,
\be
\delta {\mathfrak s} = \frac{1}{\rho T} \left( \delta \epsilon_0 - \frac{\epsilon_0+p}{\rho}\, \delta \rho\right) ,
\ee
$\epsilon_0$ being the value of the energy density $\epsilon$ in a frame where ${\bf v}=0$.
Combining the linearized equations~(\ref{hydro-eq-rho}) and~(\ref{hydro-eq-eps}), we obtain the equation for the local deviations of the specific entropy:
\be
\rho\, T \, \partial_t \, {\mathfrak s} = - \nabla^a {\cal J}_{\epsilon}^a \, , \label{macro-eq-s}
\ee
which shows that the time evolution would be adiabatic, i.e., isoentropic, if transport was negligible.

Therefore, we consider the following equivalent equations,
\begin{align}
& \partial_t \, y = -\frac{\xi^{\prime}}{T} \, \nabla^2 T - \zeta \, \nabla^a \nabla^b \phi^{ab} + \zeta \, \mu_{y,\boldsymbol{\phi}} \, \nabla^2 \phi^{aa} + {\cal D}_y \, \nabla^2 y \, , \label{macro-eq-y-2}\\
& \rho\, T \, \partial_t \, {\mathfrak s} = \kappa^{\prime} \, \nabla^2 T + \xi \, \nabla^a \nabla^b \phi^{ab} - \xi \, \mu_{y,\boldsymbol{\phi}} \, \nabla^2\phi^{aa} - {\cal K}_y \, \nabla^2 y \, , \label{macro-eq-s-2}\\
& \rho\, \partial_t \, v^b = \nabla^a \sigma^{ab}  + \eta^{abcd}\, \nabla^a\nabla^c v^d \, , \label{macro-eq-v-2}\\
& \partial_t \, u^b = v^b +\frac{\xi^{\prime}}{T} \, \nabla^b T + \zeta \, \nabla^a \phi^{ab} - \zeta \, \mu_{y,\boldsymbol{\phi}} \, \nabla^b \phi^{aa}- {\cal D}_{y} \, \nabla^b y \, . \label{macro-eq-u-2}
\end{align}
This set of linear equations can be closed and solved with perturbative expansion following the same lines as in Ref.~\cite{MG21}.  For this purpose, we expand the deviations of the fields $X=({\mathfrak s},p,\phi^{ab},\sigma^{ab})$ in terms of the deviations of $(u^{ab},T,y)$ according to
\begin{align}
\delta X = \left(\frac{\partial X}{\partial u^{cd}}\right)_{\boldsymbol{\mathsf u^\prime},T,y} \delta u^{cd} &+ \left(\frac{\partial X}{\partial T}\right)_{\boldsymbol{\mathsf u},y} \delta T+ \left(\frac{\partial X}{\partial y}\right)_{\boldsymbol{\mathsf u},T} \delta y \, . 
\label{dX}
\end{align}

The elastic properties of the crystal are characterized by the isothermal stress-strain coefficients~\cite{W98}
\be
B_T^{abcd} \equiv \left(\frac{\partial \sigma^{ab}}{\partial u^{cd}}\right)_{\boldsymbol{\mathsf u^\prime},T,y} 
\ee
and
\be
\label{G-dfn}
G_T^{abcd} \equiv \left(\frac{\partial \phi^{ab}}{\partial u^{cd}}\right)_{\boldsymbol{\mathsf u^\prime},T,y} \, .
\ee
For cubic crystals, rank-four tensors can be expressed in terms of only three coefficients because of lattice symmetries.  Moreover, rank-two tensors are diagonal in cubic crystals and we thus have that
\be
\left(\frac{\partial p}{\partial u^{ab}}\right)_{\boldsymbol{\mathsf u^\prime},T,y} = -B_T \, \delta^{ab}
\ee
with the isothermal bulk modulus
\be\label{B_T}
B_T \equiv \rho \left(\frac{\partial p}{\partial \rho}\right)_{T,y} .
\ee
Consequently, we here obtain the relation
\be
B_T^{abcd} = B_T \, \delta^{ab} \, \delta^{cd} + G_T^{abcd} \, . \label{sigma-p-phi-u-cubic}
\ee
In the equilibrium rest state of a cubic crystal simulated by periodic boundary conditions, the stress tensor is diagonal giving the hydrostatic pressure according to $\sigma^{ab}\vert_{\rm eq}= -p \delta^{ab}$, so that $\phi^{ab}\vert_{\rm eq}=0$.  As a consequence, the following equations hold,
\be
\left(\frac{\partial \phi^{ab}}{\partial T}\right)_{\boldsymbol{\mathsf u},y} = 0 \, ,
\qquad
\left(\frac{\partial \phi^{ab}}{\partial y}\right)_{\boldsymbol{\mathsf u},T} = 0 \, .
\ee
Moreover, in cubic crystals, the coefficients~(\ref{G-dfn}) satisfy the identity
\be
G_T^{aacd} = 0 \, .
\ee
Because of the relation of the triple product $\left(\frac{\partial p}{\partial T}\right)_{\rho,y}=-\left(\frac{\partial p}{\partial \rho}\right)_{T,y}\left(\frac{\partial\rho}{\partial T}\right)_{p,y}$, we furthermore have that
\be
\label{dsigma/dT}
\left(\frac{\partial \sigma^{ab}}{\partial T}\right)_{\boldsymbol{\mathsf u},y}  = - \left(\frac{\partial p}{\partial T}\right)_{\rho,y} \, \delta^{ab} = - \alpha \, B_T \, \delta^{ab}
\ee
with the isothermal expansion coefficient
\be
\alpha \equiv - \frac{1}{\rho} \left(\frac{\partial\rho}{\partial T}\right)_{p,y} \, .
\ee
In addition, we define the derivative of the pressure with respect to the vacancy molar fraction as
\be
\label{dpdy}
\pi_y \equiv  \left(\frac{\partial p}{\partial y}\right)_{\boldsymbol{\mathsf u},T} ,
\ee
implying
\be
\left(\frac{\partial \sigma^{ab}}{\partial y}\right)_{\boldsymbol{\mathsf u},T} = -\pi_y \, \delta^{ab} \, .
\ee

As a consequence of the Maxwell relation
\be
\left(\frac{\partial{\mathfrak s}}{\partial u^{ab}}\right)_{\boldsymbol{\mathsf u^\prime},T,y} = - \frac{1}{\rho} \left(\frac{\partial \sigma^{ab}}{\partial T}\right)_{\boldsymbol{\mathsf u},y}
\ee
(see Ref.~\cite{W98}) and Eq.~(\ref{dsigma/dT}), we find that
\be
\left(\frac{\partial{\mathfrak s}}{\partial u^{ab}}\right)_{\boldsymbol{\mathsf u^\prime},T,y} = \frac{\alpha B_T}{\rho} \, \delta^{ab} \, .
\ee
Moreover, we have that
\be
\left(\frac{\partial{\mathfrak s}}{\partial T}\right)_{\boldsymbol{\mathsf u},y} = \frac{c_v}{T}
\ee
in terms of the specific heat capacity at constant volume $c_v$.  We also need to introduce the derivative of the specific entropy with respect to the vacancy molar fraction according to
\be\label{dsdy}
\varsigma_y \equiv \left(\frac{\partial{\mathfrak s}}{\partial y}\right)_{\boldsymbol{\mathsf u},T} \, .
\ee

For cubic crystals, we thus have
\begin{align}
\delta {\mathfrak s} &= \frac{\alpha\, B_T}{\rho}\, \delta u^{aa} + \frac{c_v}{T}\, \delta T + \varsigma_y \, \delta y \, , \label{s-uTy-cubic}\\
\delta p &= -B_T \, \delta u^{aa} + \alpha \, B_T \, \delta T + \pi_y \, \delta y \, , \label{p-uTy-cubic}\\
\delta \phi^{ab} &= G_T^{abcd} \, \delta u^{cd} \, , \label{phi-uTy-cubic}\\
\delta \sigma^{ab} &= B_T^{abcd} \,  \delta u^{cd} - \alpha B_T \, \delta^{ab} \, \delta T - \pi_y \, \delta^{ab} \, \delta y \, . \label{sigma-uTy-cubic} \qquad
\end{align}
Equation~(\ref{s-uTy-cubic}) can be used to express $\delta T$ as a linear combination of $(\delta y,\delta\mathfrak{s},\delta u^{aa})$ and, thus, to eliminate $\delta T$ from the right-hand sides of Eqs.~(\ref{macro-eq-y-2})-(\ref{macro-eq-u-2}).

\subsection{Closed set of linearized equations}

Using, Eqs.~(\ref{s-uTy-cubic})-(\ref{sigma-uTy-cubic}), the evolution equations~(\ref{macro-eq-y-2})-(\ref{macro-eq-u-2}) can be transformed into the following closed set of linear equations for the deviations of the eight fields $(y,\mathfrak{s},v^a,u^a)$ around the equilibrium rest state:
\begin{align}
\partial_t  \, \delta y = & \ D_{yy} \, \nabla^2 \delta y + D_{y{\mathfrak s}} \, \nabla^2 \delta{\mathfrak s} + D_{yu}^{abcd} \, \nabla^a \nabla^b \nabla^c \delta u^d , \qquad\ \label{macro-eq-y-fin}\\
\partial_t  \, \delta {\mathfrak s} =&\  D_{{\mathfrak s}y} \, \nabla^2 \delta y + D_{\mathfrak{ss}} \, \nabla^2 \delta {\mathfrak s} + D_{{\mathfrak s}u}^{abcd} \, \nabla^a \nabla^b \nabla^c \delta u^d , \label{macro-eq-s-fin}\\
\partial_t \, \delta v^a =&\ C_{vy} \, \nabla^a \delta y + C_{v{\mathfrak s}} \, \nabla^a \delta{\mathfrak s}  + D_{vv}^{abcd}\, \nabla^b\nabla^c \delta v^d \nonumber\\
& + C_{vu}^{abcd}\, \nabla^b\nabla^c \delta u^d , \label{macro-eq-v-fin}\\
\partial_t \, \delta u^a =&\ D_{uy} \, \nabla^a \delta y + D_{u{\mathfrak s}} \, \nabla^a \delta{\mathfrak s}  +  \delta v^a + D_{uu}^{abcd}\, \nabla^b \nabla^c \delta u^d , \label{macro-eq-u-fin}
\end{align}
with the coefficients
\begin{align}
D_{yy} &\equiv {\cal D}_y  + \frac{\varsigma_y}{c_v} \, \xi^{\prime} \, , \qquad\label{D_yy}\\
D_{y{\mathfrak s}} &\equiv -\frac{1}{c_v} \, \xi^{\prime}  \, , \qquad \label{D_ys}\\
D_{yu}^{abcd} &\equiv - G_T^{abcd} \, \zeta + \frac{\alpha\, B_T}{\rho c_v} \, \xi^{\prime} \, \delta^{ab} \, \delta^{cd} \, , \qquad \label{D_yu}\\
D_{{\mathfrak s}y} &\equiv -\frac{\varsigma_y}{\rho c_v}\, \kappa^{\prime} - \frac{1}{\rho T} \, {\cal K}_y \, , \label{D_sy}\\
D_{\mathfrak{ss}} &\equiv \frac{1}{\rho c_v} \, \kappa^{\prime} \, , \label{D_ss}\\
D_{{\mathfrak s}u}^{abcd} &\equiv \frac{1}{\rho T}\, G_T^{abcd} \, \xi  - \frac{\alpha\, B_T}{\rho^2 c_v} \, \kappa^{\prime} \,  \delta^{ab} \, \delta^{cd} \, , \label{D_su}\\
C_{vy} &\equiv \frac{1}{\rho} \left(- \pi_y + \varsigma_y \frac{\alpha\, B_T T}{c_v} \right) \, , \label{C_vy}\\
C_{v{\mathfrak s}} &\equiv - \frac{\alpha B_T T}{\rho c_v} \, , \label{C_vs}\\
D_{vv}^{abcd} &\equiv \frac{\eta^{abcd}}{\rho} \, , \label{D_vv}\\
C_{vu}^{abcd} &\equiv \frac{B_{\mathfrak s}^{abcd}}{\rho} \, , \qquad \label{C_vu}\\
D_{uy}  &= - D_{yy} \, , \quad D_{u{\mathfrak s}} = - D_{y{\mathfrak s}} \,  , \quad D_{uu}^{abcd}  = - D_{yu}^{abcd} \, , \qquad\  \label{D_uy-D_us-D_uu} 
\end{align}
where $\kappa^{\prime}$, $\xi^{\prime}$, ${\cal K}_y$, and ${\cal D}_y$ are respectively defined by Eqs.~(\ref{kappa-prime}), (\ref{xi-prime}), (\ref{K_y}), and~(\ref{D_y}), and
\be
\label{B_s}
B_{\mathfrak s}^{abcd} \equiv B_T^{abcd}  + (\gamma-1) B_T \, \delta^{ab} \, \delta^{cd}
\ee
are the adiabatic (isoentropic) stress-strain coefficients, $\gamma\equiv c_p/c_v$ being the specific heat ratio.  The relation $c_p-c_v=T\alpha^2 B_T/\rho$ has been used in the calculations.
We note that the relations~(\ref{D_uy-D_us-D_uu}) are the consequence of the fact that ${\cal J}_y^a=-{\cal J}_{u^a}$.


\section{Hydrodynamic modes}
\label{sec:hydro-modes}

The hydrodynamic modes are the solutions of Eqs.~(\ref{macro-eq-y-fin})-(\ref{macro-eq-u-fin}) with the spatial dependence $\exp(-{\rm i}{\bf q}\cdot{\bf r})$ of given wave vector $\bf q$.  These equations can be decoupled if the wave vector $\bf q$ is taken in one of the special directions $[100]$, $[110]$, and $[111]$ of the cubic lattice.  The calculation of the hydrodynamic modes,  summarized in App.~\ref{app:hydro-modes}, gives the dispersion relations for the eight modes of the crystal, including the six longitudinal and transverse sound modes, the heat mode, and the vacancy diffusion mode.  These dispersion relations determine the frequencies and the widths of the resonances of the hydrodynamic spectral functions~\cite{F75,BP76,BY80,MG24a,MG24b}.  The dispersion relations depend not only on the magnitude $q=\Vert{\bf q}\Vert$ of the wave vector, but also on its direction ${\bf q}/q\in\{[100],[110],[111]\}$.  The macroscopic hydrodynamic theory gives the dispersion relations up to the terms of $O(q^2)$.  Beyond  the quadratic order in the wave number $q$, the long-time tail effects may contribute by corrections of $O(q^{5/2})$ \cite{DvBK21}.  Here, the main focus is on the eighth mode, i.e., the mode of  vacancy diffusion.

\subsection{Sound modes}

The six sound modes are propagative with dispersion relations of the form $z_\alpha=\pm{\rm i} c_\sigma q - \Gamma_\sigma q^2$ in terms of the wave number $q$ of the modes $\alpha=1$-$6$ with $\sigma={\rm l},{\rm t}_1,{\rm t}_2$.  The longitudinal speed of sound $c_{\rm l}$ is given by Eq.~(\ref{A-l}) and the transverse speeds of sound $c_{{\rm t}_1}$ and $c_{{\rm t}_2}$ by Eq.~(\ref{A-t}).  The speeds of sound have the values expected from previous work~\cite{MG24b}.  The acoustic attenuation coefficients $\Gamma_\sigma$ are expressed by Eq.~(\ref{z1_123+456}) in terms of the transport properties, including the effects of vacancy conduction $\zeta$ and thermodiffusion $\xi$.

\subsection{Heat conduction}

The dispersion relation of the heat mode is diffusive and non-propagative of the form $z_7=-\chi\, q^2$ according to Eqs.~(\ref{z0_7}) and~(\ref{z1_7}).  Since $z_7^{(0)}=0$, the diffusivity of heat conduction is given by the factor in front of $q^2$ in the right-hand side of Eq.~(\ref{z1_7}).  Substituting therein Eq.~(\ref{D_ss}) for $D_{\mathfrak{ss}}$, Eq.~(\ref{C_vs}) for $C_{v\mathfrak{s}}$, the components $D_{\mathfrak{s}u}^{\rm ll}$ of~(\ref{D_su}) given by Eq.~(\ref{A2-D2}) with $\sigma=\sigma'={\rm l}$, and using Eq.~(\ref{A-l}), we obtain the following expression for the diffusivity of heat conduction,
\be\label{chi-final}
\chi = \frac{\kappa^{\prime}}{\rho c_v} \, \frac{B_T^{\rm l}}{B_T^{\rm l}+(\gamma -1) B_T}
+ \frac{\xi}{\rho c_v} \, \frac{\alpha B_T\left(B_T^{\rm l} - B_T\right)}{B_T^{\rm l}+(\gamma -1) B_T}
\ee
with $\kappa^{\prime}$ given by Eq.~(\ref{kappa-prime}).
In the limit of a perfect crystal where vacancy thermodiffusion is absent ($\xi=0$), we recover the result
\be\label{chi-pft-crystal}
\chi = \frac{\gamma D_T}{1+(\gamma-1)\frac{B_T}{B_T^{\rm l}}}
\qquad\mbox{with}\qquad
D_T \equiv \frac{\kappa}{\rho c_p} \, ,
\ee
as previously obtained in Ref.~\cite{MG24b} (where we used the notation $B^T_{\rm l}=B_T^{\rm l}$).

We note that Eq.~(\ref{chi-final}) could be further simplified using $B_T^{\rm l}+(\gamma -1) B_T=\rho c_{\rm l}^2$ in terms of the speed $c_{\rm l}$ of the longitudinal sound waves.  We point out that the second term of Eq.~(\ref{chi-final}) is proportional to $\xi$ (and not $\xi^{\prime}$).  Since the coefficient $B_T^{\rm l}$ and the speed $c_{\rm l}$ depend on the direction of the wave vector $\bf q$, this is also the case for the heat diffusivity~(\ref{chi-final}).

\subsection{Vacancy diffusion}

The dispersion relation of the vacancy diffusion mode is also diffusive and non-propagative of the form $z_8=-D_{\rm vac}\, q^2$ according to Eqs.~(\ref{z0_8}) and~(\ref{z1_8}).  Since $z_8^{(0)}=0$, the diffusivity of the vacancy diffusion mode is given by the factor in front of $q^2$ in the right-hand side of Eq.~(\ref{z1_8}).  Substituting therein Eq.~(\ref{D_yy}) for $D_{yy}$, Eq.~(\ref{C_vy}) for $C_{vy}$, the components $D_{yu}^{\rm ll}$ of~(\ref{D_yu}) given by Eq.~(\ref{A2-D2}) with $\sigma=\sigma'={\rm l}$, and using Eq.~(\ref{A-l}), we find the following expression for the diffusivity of the eighth mode,
\be
D_{\rm vac} = {\cal D}_y + \Delta D_{\rm vac}
\label{D_vac-final-1}
\ee
with
\begin{align}
\Delta D_{\rm vac} = &\ \frac{\varsigma_y}{c_v} \, \xi^{\prime} \nonumber\\
&+ \frac{-\pi_y + \varsigma_y \, \alpha B_T T/c_v}{B_{T}^{\rm l} + (\gamma-1) B_T} \Big[ \big( B_T^{\rm l} - B_T \big) \zeta  - \frac{\alpha B_T}{\rho c_v} \, \xi^{\prime} \Big] , 
\label{D_vac-final-2}
\end{align}
where ${\cal D}_y$ is given by Eq.~(\ref{D_y}) and $\xi^{\prime}$ by Eq.~(\ref{xi-prime}).

For cubic crystals, the coefficient ${\cal D}_y$ is an isotropic scalar property, but the correction $\Delta D_{\rm vac}$ depends on the direction of the wave vector $\bf q$ because of the presence of the coefficient $B_T^{\rm l}$ in Eq.~(\ref{D_vac-final-2}).   This dependence results from the coupling of the motion of vacancies to the sound modes.  This mode-coupling effect induces a collective motion of the vacancies. As shown in App.~\ref{app:S_vac}, the diffusivity $D_{\rm vac}$ determines the poles of the vacancy spectral function.

\subsection{Limit of low vacancy density}
\label{ssec:llvd}

The reasoning based on the local conservation equation of vacancies~(\ref{hydro-eq-y-2}) and the expression~(\ref{drift-velocity-y}) for the vacancy driving velocity shows that the coefficients of vacancy conduction $\zeta$ and thermodiffusion $\xi$ are proportional to the vacancy molar fraction $y$.  Therefore, they vanish as $\zeta=O(y)$ and $\xi=O(y)$ in the limit $y\to 0$ of arbitrarily low vacancy density.

In this limit, we recover the dispersion relations of the seven hydrodynamic modes of the perfect crystal, as obtained in Ref.~\cite{MG24b}.  Indeed, Eq.~(\ref{z1_123+456}) with the coefficients in Eqs.~(\ref{D_yu}), (\ref{D_su}), (\ref{D_vv}), and~(\ref{D_uy-D_us-D_uu}) shows that the acoustic attenuation coefficients $\Gamma_\sigma$ converge to their perfect-crystal values of Ref.~\cite{MG24b} in the limit $y\to 0$.  Similarly, the dispersion relation~(\ref{chi-final}) for heat conduction also converges to its perfect-crystal value~(\ref{chi-pft-crystal}) for the same reason that $\zeta=0$ and $\xi=0$ if $y=0$.   Because the contributions of vacancy conduction and thermodiffusion vanish in this limit,  the acoustic attenuation coefficients and the diffusivity of heat conduction thus recover their expected dependence on the heat conductivity $\kappa$ and the viscosity coefficients $\eta_{11}$, $\eta_{12}$, and $\eta_{44}$ \cite{MG24b}.

In the presence of vacancies, the eighth mode has a dispersion relation with the diffusivity~(\ref{D_vac-final-1})-(\ref{D_vac-final-2}). This coefficient now remains nontrivial in the limit $y\to 0$ because of the coefficient ${\cal D}_y$, which would be missing if the chemical potential of vacancies was not explicitly taken into account.  Indeed, in the limit $y\to 0$ where the vacancies form a dilute system, the chemical potential of the vacancies is known to behave as
\be
\mu_y = \mu_y^0(T,\boldsymbol{\phi}) + n_{\rm eq,0} \, k_{\rm B} T \, \ln y  + O(y) \, ,
\ee
where $\mu_y^0(T,\boldsymbol{\phi})=n_{\rm eq,0} g_{\rm v}$ is the reference value calibrated for $y=1$ according to Eq.~(\ref{mu_v}) and $\boldsymbol{\phi}=(\phi^{ab})$ \cite{HL64,AL87,LC85,PF01}.  Consequently, the vacancy diffusion coefficient is given at leading order in $y$ by
\be
\label{D-zeta}
{\cal D}_y \equiv \left(\frac{\partial\mu_y}{\partial y}\right)_{T,\boldsymbol{\phi}} \zeta = n_{\rm eq,0} \, \frac{k_{\rm B}T}{y} \, \zeta \, ,
\ee
which is finite and positive in the limit $y\to 0$ since $\zeta=O(y)$.  The corrective terms given by Eq.~(\ref{D_vac-final-2}) are all equal to zero for $y=0$.  Indeed, the derivative of the specific entropy with respect to $y$ behaves as $\varsigma_y=O(\ln y)$ according to Eq.~(\ref{dSdy/Na-HS}), but it is multiplied by $\xi^{\prime}=O(y)$ and $\zeta=O(y)$.  Moreover, the derivative of the pressure with respect to $y$ is constant in the limit $y\to 0$, but it is also multiplied by $\xi^{\prime}=O(y)$ and $\zeta=O(y)$.  Therefore, the diffusivity of the vacancy mode goes as
\be
D_{\rm vac} = {\cal D}_y + O(y\ln y) \, ,
\ee
which is now finite, positive, and taking a physically expected value.

Moreover, we now have consistency with the value expected with the Markovian jump model for the random walk of a vacancy in the crystal (see App.~\ref{app:Markov}).  If $a$ denotes the size of the cubic cells of the face-centered cubic (fcc) crystal and $w$ is the jump rate, the coefficient given by Eq.~(\ref{zeta-H-y}) with the Helfand moment~(\ref{Helfand-y}) for $N_{\rm v}=1$ can be approximated by
\be
\zeta \simeq \frac{y}{n_{\rm eq,0} \, k_{\rm B}T} \, \frac{w\, a^2}{12} \, ,
\ee
so that the formula~(\ref{D-zeta}) gives
\be
\label{eq:D_omega}
{\cal D}_y  \simeq \frac{w\, a^2}{12} \, ,
\ee
which is precisely the value expected for the diffusion coefficient of the Markovian jump model in a fcc crystal.

In the hard-sphere crystal, the vacancy chemical potential is proportional to the temperature, so that $\big[\partial_T(\mu_y/T)\big]_{y,\boldsymbol{\phi}}=0$, in which case we have that $\kappa^{\prime}=\kappa$ and $\xi^{\prime}=\xi$ by Eqs.~(\ref{kappa-prime}) and~(\ref{xi-prime}).

The conclusion is that, for consistency, it is essential to include the chemical potential   of the vacancies, otherwise, the genuine diffusion coefficient is missing in the macroscopic equations for the hydrodynamics of  nonperfect crystals.


\section{The hard-sphere crystal}
\label{sec:num_res_HS}

Based on the predictions of the hydrodynamic theory of solids, we study the vacancy diffusion for  a crystal of  hard spheres using numerical methods. We first calculate the transport coefficients $\zeta$ and $\xi$ associated with the conduction and the thermodiffusion of vacancies, and the diffusion coefficient ${\cal D}_y$. Next, Eqs.~(\ref{D_vac-final-1})-\ref{D_vac-final-2}) are used to predict the dispersion relation of the eighth mode of vacancy diffusion.

\subsection{Hard-sphere dynamics}
\label{ssec:HS_dym}

The hard-sphere system consists of $N$ identical spheres of mass $m$ and diameter $d$ evolving in a simulation box of volume $V$ with periodic boundary conditions. The edges of this box are along the $x$, $y$, and $z$ axes and are of lengths $L_x$, $L_y$, and $L_z$ such that $V=L_xL_yL_z$. The position and the momentum of the $i^{\rm th}$ particle are ${\bf r}_i$ and  ${\bf p}_i$ respectively. The spheres interact by a pair potential that vanishes if the interparticle distance $r_{ij}$, computed with the minimum image convention~\cite{H97}, is larger than $d$ and is otherwise infinite. In the absence of overlap between the particles, the Hamiltonian of the system reduces to the kinetic energy, which is conserved, and the ensemble of the  simulation is $(N,V,E)$. The total momentum ${\bf P}=\sum_{i=1}^N{\bf p}_i$ is also conserved and the temperature is related to the total energy by $k_{\rm B} T = (2/3)(E/N)$.

The simulation is performed with an event-driven algorithm~\cite{H97}. The particles evolve in free flights interrupted by instantaneous elastic collisions at times $\{t_c\}$. During a free flight, the position of the $i^{\rm th}$ particle is given by ${\bf r}_i(t)={\bf r}_i(t_c)+{\bf p}_i(t_c+0)(t-t_c)/m$ with $t\in[t_c,t_{c+1})$. At a collision, two particles exchange a momentum $\Delta {\bf p}^{(c)}_{ij}=-({2}/{d^2})[{\bf r}^{(c)}_{ij}\cdot{\bf p}^{(c)}_{ij}(t_c-0)]{\bf r}^{(c)}_{ij}$ expressed in terms of the canonically conjugate positions ${\bf r}_{ij}\equiv{\bf r}_i-{\bf r}_j$ and momenta ${\bf p}_{ij}\equiv({\bf p}_i-{\bf p}_j)/2$. The algorithm also updates the list of the future collisions by solving the quadratic equations $[{\bf r}_{i_c}(t)-{\bf r}_k(t)]^2=d^2$ for $t  >t_c$, where $i_c$ labels the two particles colliding at time $t_c$ and $k$ all the other particles. 

 In the simulation, the diameters and the masses of the spheres are set equal to $1$. The initial momenta of the particles are sampled from a Gaussian distribution with the constraints that ${\bf P}=0$ and that the temperature $k_{\rm B}T=1$. The trajectories of the hard spheres are generated by the algorithm, giving their positions  and momenta at any time $t$ of the simulation.  Data are collected after a transient evolution of time $t_{\rm transient}$. The ensemble averages of any quantity $X$ are obtained by considering $N_{\rm stat}$ trajectories of $n_{\rm step}$ time steps with increment $\Delta t$ such that $\langle X\rangle_{\rm eq}=N_{\rm stat}^{-1}\sum_{k=1}^{N_{\rm stat}}X^{(k)}$. In the following, the results are expressed in terms of dimensionless quantities denoted by an asterisk.  The dimensionless particle density is defined as $n_{0*} \equiv n_0d^3 = N_0d^3/V$. The dimensionless positions, momenta, and time are given by $ {\bf r}_{i*} \equiv   {\bf r}_{i}/d$, ${\bf p}_{i*} \equiv {\bf p}_{i}/ \sqrt{mk_{\rm B}T}$, and $t_* \equiv (t/d) \sqrt{k_{\rm B}T /m}$. The Helfand moments, the thermodynamic quantities, and the transport and diffusion coefficients read
\begin{align}
{\mathbb G}_{\epsilon}^a &= dk_{\rm B}T \, {\mathbb G}_{\epsilon *}^a \, , &  {\mathbb G}_{y}^a &= d^4\ {\mathbb G}_{y *}^a, \nonumber\\
 p &= \frac{k_{\rm B}T}{d^3} \, p_* \, , & B_{T}&=\frac{k_{\rm B}T}{d^3}B_{T*}\,,\nonumber\\c_{p}&=\frac{k_{\rm B}}{m}\ c_{p*}\;,
 & \kappa^{ab} &= \frac{k_{\rm B}}{d^2}\sqrt{\frac{k_{\rm B}T}{m}} \, \kappa^{ab}_*\;,\nonumber\\
 \zeta^{ab}&= \frac{d^4}{\sqrt{m k_{\rm B}T}}\ \zeta^{ab}_*\;,&\xi^{ab}&= {d}{\sqrt{\frac{k_{\rm B}T}{m}}}\ \xi^{ab}_*\;,\nonumber\\
 {\cal D}_y & =  d\sqrt{\frac{k_{\rm B}T}{m}} {\cal D}_{y*} \, , &  {D}_{\rm vac} & =  d\sqrt{\frac{k_{\rm B}T}{m}} {D}_{\rm vac *} \, , 
\end{align}
in terms of the corresponding dimensionless quantities.

\subsection{Simulation of a hard-sphere crystal with vacancies}
\label{ssec:NPHSC}

The  spheres are initially placed on a fcc lattice composed of cubic cells of edge $a$. In each direction, there are $M_{x}$, $M_{y}$, and $M_{z}$ cubic cells. In each cell, the lattice has four nodes located at the positions ${\bf R}_j =x_j {\bf e}_x +y_j {\bf e}_y + z_j {\bf e}_z$ with $j\in\{1,2,3,4\}$ and $a^{-1}(x_j,y_j,z_j)\in\{(\frac{1}{4},\frac{1}{4},\frac{1}{4}),(\frac{3}{4},\frac{3}{4},\frac{1}{4}),(\frac{3}{4},\frac{1}{4},\frac{3}{4}),(\frac{1}{4},\frac{3}{4},\frac{3}{4})\}$~\cite{AM76} and the total number of lattice sites is $N_0=4M_{x}M_{y}M_{z}$. For a perfect crystal, all  the sites of the lattice are occupied by exactly one particle. For a nonperfect crystal,  a number $N_{\rm v}$ of lattice sites are left unoccupied, such that $N+N_{\rm v}=N_0$. The distance between the vacancies is initially maximized when $N_{\rm v}>1$. The numbers $N_0$, $N$, and $N_{\rm v}$ remain constant during the simulation. The equilibrium density of the lattice is given by $n_0=4/a^3=N_0/V$. The simulation is performed for densities between the melting density $n_{\rm m*}=1.037 \pm 0.003$~\cite{S98,PBHB20} and the close packing density $n_{\rm cp*}=\sqrt{2}$, where the hard-sphere crystal exists.

The positions of the vacancies are accessible at any time of the simulation. The method consists in finding the closest lattice site to each particle $i$ by solving ${\rm min}_{j}[{\bf r}_i(t)-{\bf R}_j(t)]^2$, where ${\bf R}_j(t)$ is the position of the $j^{\rm th}$ lattice node at time $t$. The positions of the vacancies ${\bf r}_{{\rm v},i}$ are given by the locations of the lattice sites that are not assigned to any particle. Note that the vacancies are always located with respect to the current position of the lattice. Indeed,   each jump of the vacancy is echoed with a jump of a particle to a vacant neighboring site. If the particle jumps to a site at a distance $a/2$ in a given direction, the lattice drifts by a distance $a/(2N)$ in the opposite direction, due to the conservation of the center of mass position. The position of a lattice site at time $t$ is thus ${\bf R}_j(t)={\bf R}_j(0)+\Delta {\bf R}(t)$, where $\Delta {\bf R}(t)$ is the drift of the lattice. 

There is no event where a vacancy-interstitial pair is generated in the numerical simulations, indicating that the rate of such pair creation is negligible in the conditions considered, which justifies the assumption of conservation of the number of vacancies.

\subsection{Helfand moments for the hard-sphere dynamics}
\label{ssec:Helfand-HS}

According to microscopic hydrodynamics, the transport coefficients are given in terms of the covariances of the Helfand moments. We compute the vacancy conductivity $\zeta$ and the coefficient $\xi$ of vacancy thermodiffusion using the Einstein-Helfand formulas given in Eqs.~\eqref{xi-H-y}-\eqref{zeta-H-y} and following the method  described in Ref.~\cite{MG24a}. The Helfand moment associated with the transport of energy is known to read
\begin{align}
{\mathbb G}_{\epsilon}^a(t)=&\sum_{\substack{(c\to c+1)\\\,\in\,[0,t]}} \sum_{i} \frac{{p}^a_i(t_c+0) }{m}\, \frac{[{\bf p}_{i}(t_c+0)]^2}{2m}\, (t_{c+1}-t_c) \nonumber\\
& + \sum_{c\, \in \, [0,t]}  r^{a (c)}_{ij} \, \Delta{p}_{ij}^{b (c)} \, \frac{{p}^{b (c)}_{i}+{p}^{b (c)}_{j}}{2m}\, ,\label{eq:GeaHS}
\end{align}
where the first and second sums are the contributions of the free flights and the collisions, respectively \cite{MG24a}. The Helfand moment associated with the transport of vacancies is new and given by
\begin{align}
{\mathbb G}^a_y(t)&=\frac{1}{n_{\rm eq, 0}}\sum_{i=1}^{N_{\rm v}} \left[ {r}_{{\rm v},i}^{a}(t)-{r}_{{\rm v},i}^{a}(0)\right] ,
\end{align}
as shown in App.~\ref{app:Gy}.

\subsection{Results of the simulation}
\label{ssub:ressim}

Statistics requires enough jumps of the vacancies. Since the number of jumps decreases with the density and to limit the computational cost of the simulations, we restrict the analysis to the densities $n_{0*}\in\{1.05,1.075,1.1,1.125\}$. Except when investigating the $y$-dependence of $\zeta$, the simulations are performed with cubic systems of $N_0\in\{32,108,256,500\}$ lattice sites, corresponding to $2,3,4$ and $5$ cubic cells in each direction, and one vacancy, such that $y=1/N_0$.

\subsubsection{Transport coefficients $\zeta$ and $\xi$}
\label{ssec:tc}


\begin{figure*}[h!]\centering
{\includegraphics[width=.75\textwidth]{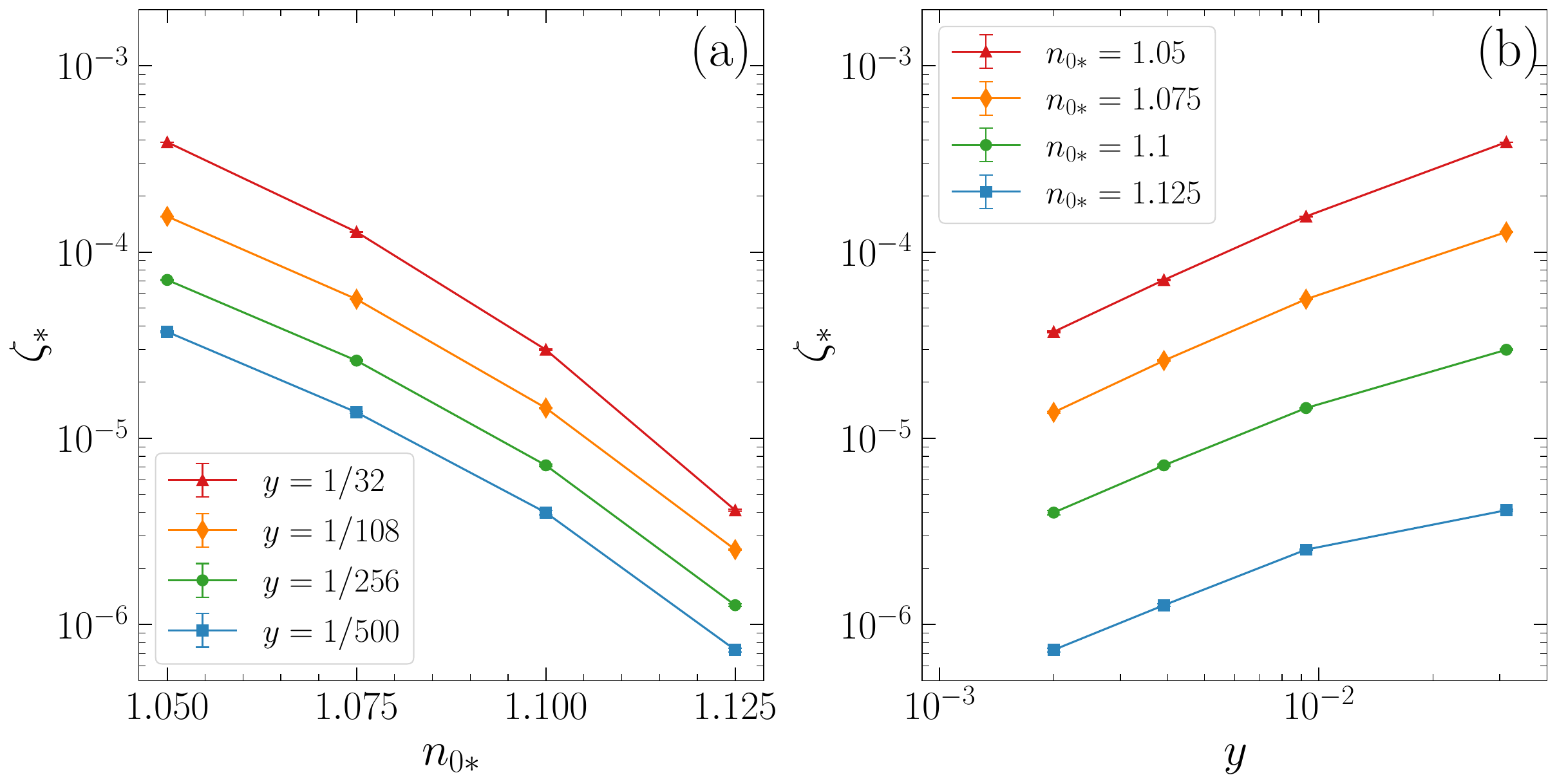}}
\caption[] {Vacancy conductivity $\zeta$ versus (a) the density $n_{0}$ and (b) the molar fraction of vacancy $y$. The  values of the coefficient~$\zeta$ are given in Table~\ref{Tab:zeta_H}.}\label{Fig:TC-zeta}
\end{figure*}


\begin{figure*}[h!]\centering
{\includegraphics[width=.75\textwidth]{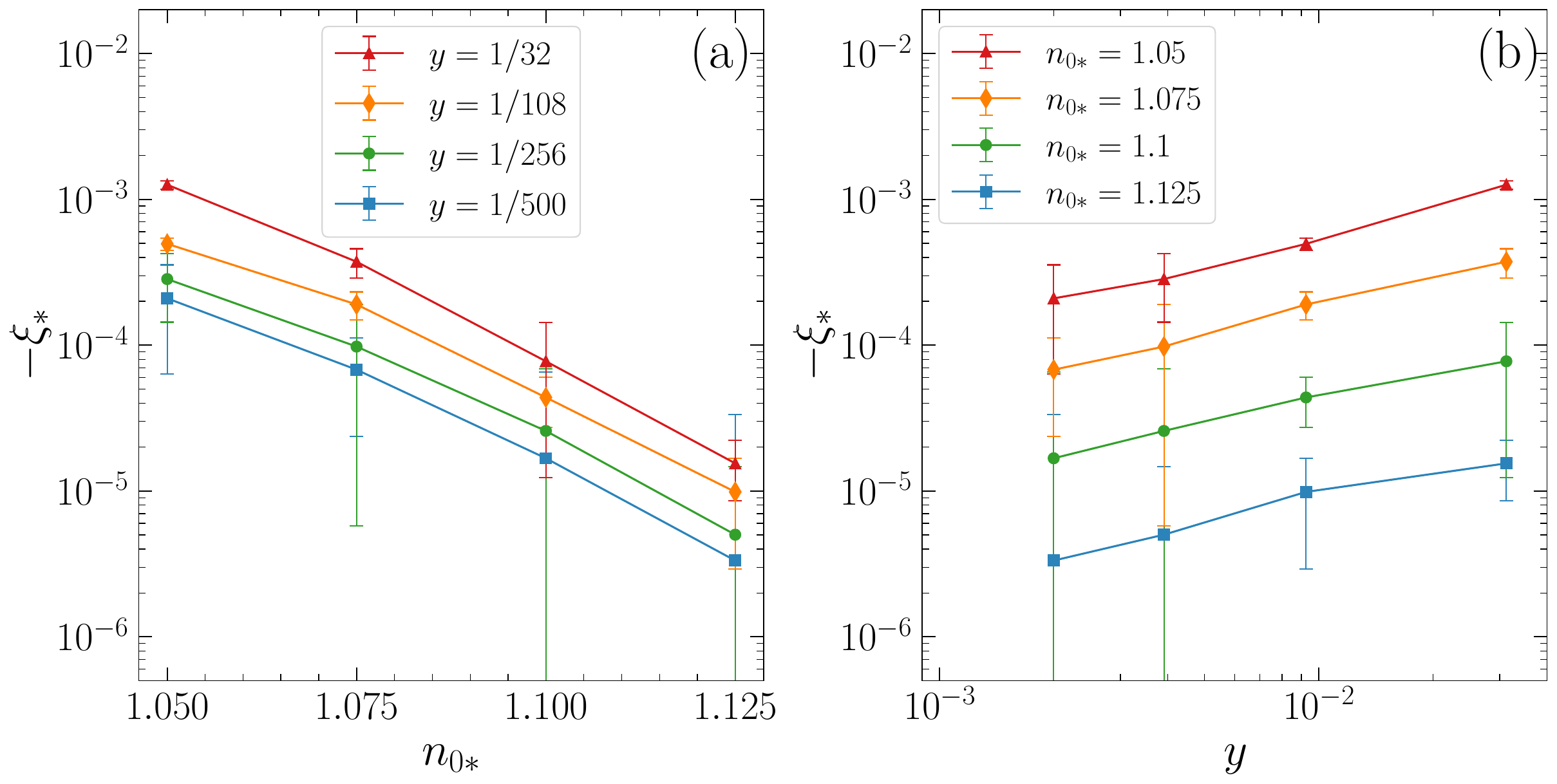}}
\caption[] {Coefficient $\xi$ of vacancy thermodiffusion versus (a) the density $n_{0}$ and (b) the molar fraction of vacancy $y$. The  values of the coefficient $\xi$ are given in Table~\ref{Tab:xi_H}.}\label{Fig:TC-xi}
\end{figure*}


The results for the transport coefficients $\zeta$ and $\xi$ are given in Tables~\ref{Tab:zeta_H}-\ref{Tab:xi_H} and shown in Figs.~\ref{Fig:TC-zeta}-\ref{Fig:TC-xi}.  Since $\xi$ includes the Helfand moment associated with the transport of energy, we report  in Table~\ref{Tab:kappa_H} the results of the heat conductivity, computed from Eq.~\eqref{kappa-H}. Note that the heat conductivity of a perfect hard-sphere crystal is given in Refs.~\cite{GAW70,MG24a,PBHB20}.  We have verified that the Helfand covariance matrices and thus the transport coefficients have the expected cubic symmetry,  as expressed in Eq.~\eqref{eq:cubsym}.

The diagonal elements of the Helfand covariance matrices have a linear dependence on time, indicating that the coefficients $\zeta$ and $\xi$ are non vanishing. Therefore, in addition to vacancy  conduction characterized by the coefficient $\zeta$, the phenomenon of vacancy thermodiffusion also exists in the hard-sphere crystal. For the thermodiffusion coefficient $\xi$, the linear dependence on time is the most explicit  for the lowest densities and largest vacancy molar fractions considered. Since this coefficient couples the transport of energy to the transport of vacancies and the Helfand moments associated with both processes differ by several orders of magnitude, as can be seen from Tables~\ref{Tab:zeta_H} and~\ref{Tab:kappa_H}, the calculation is subject to large numerical errors. The error can be reduced by increasing the number of jumps of the vacancy. Since this number decreases with both the density $n_{0}$ and the molar fraction of vacancy $y$, the result with the lowest error is obtained for $n_{0*}=1.05$ and $y=1/32$. A more systematic numerical estimation of the thermodiffusion coefficient as function of $n_{0}$ and $y$ is left for future work.

Figure~\ref{Fig:TC-zeta}(a) shows that the vacancy conductivity~$\zeta$ rapidly decreases as the crystal density $n_{0}$ increases.  Figure~\ref{Fig:TC-zeta}(b)  further shows that this coefficient increases with the vacancy molar fraction $y$.  A similar behavior is observed in Fig.~\ref{Fig:TC-xi} for minus the vacancy thermodiffusion coefficient $\xi$.   According to Eq.~(\ref{xi-H-y}), the negativity of the coefficient $\xi$ indicates that there is  a positive correlation with respect to the fluctuations of the energy Helfand moment.

\begin{figure}[t!]\centering
{\includegraphics[width=.4\textwidth]{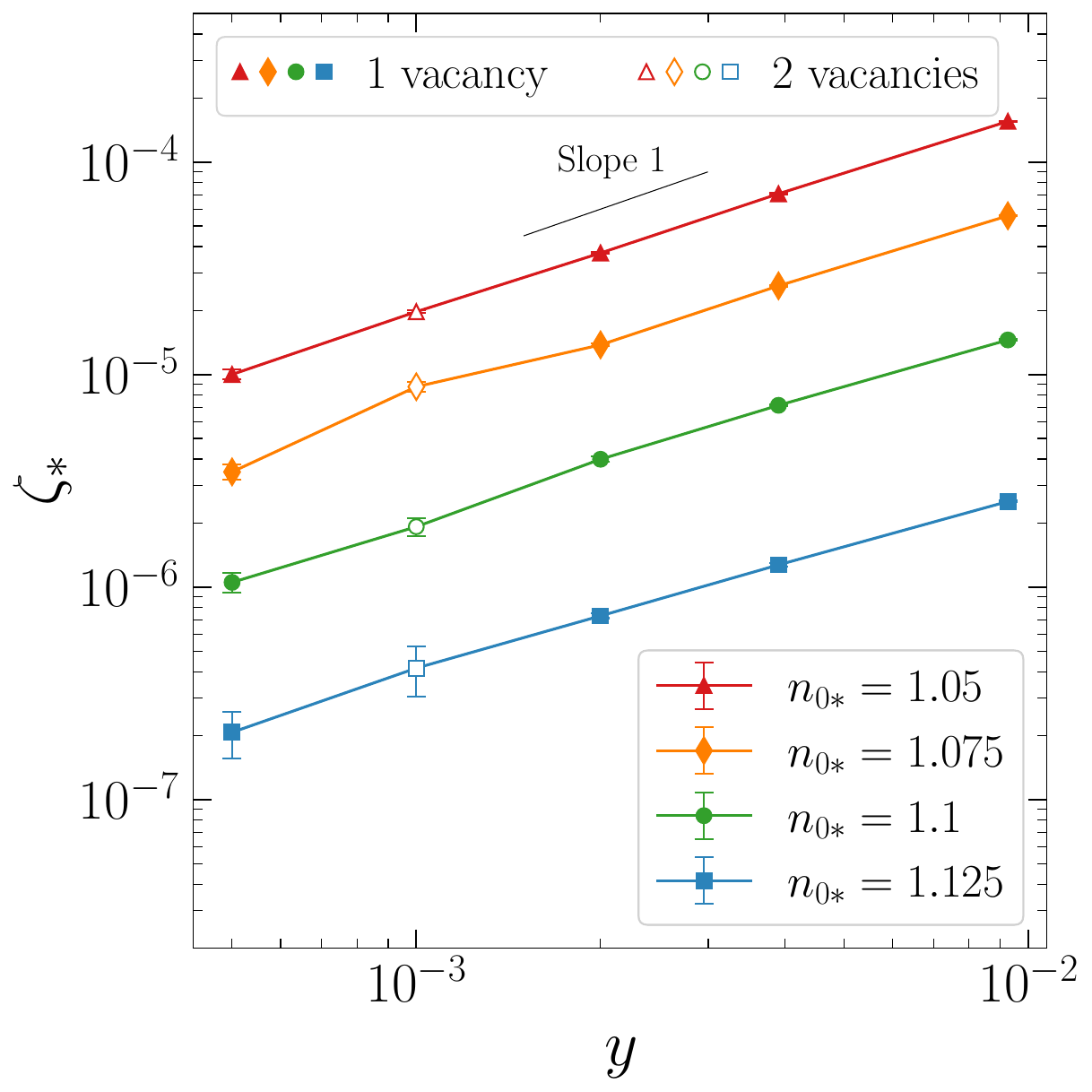}}
\caption[] {Dependence of the vacancy conductivity $\zeta$ on the molar fraction of vacancy $y$. The open symbols are computed with two vacancies in the dilute-system limit, the filled symbols with one vacancy. The data for $y\in\{1/500,1/1000,1/2000\}$ are taken from Table~\ref{Tab:zeta_y_H} and for $y\in\{1/108,1/256\}$ from Table~\ref{Tab:zeta_H}.}\label{Fig:zeta_y}
\end{figure}

We are interested in the $y$-dependence of the transport coefficients $\zeta$ and $\xi$ in the limit of vanishing $y$. The results reported in Tables~\ref{Tab:zeta_H} and~\ref{Tab:xi_H} are already pointing toward a linear dependence on $y$. A more precise analysis requires going to smaller values of $y$, hence higher $N_0$, and including multiple vacancies in the dilute-system limit. Due to the numerical errors in the computation of $\xi$ discussed  here above, we focus on $\zeta$ and consider the cases with one vacancy and $N_0\in\{500,2000\}$ and two vacancies with $N_0=2000$, such that $y\in\{1/500,1/1000, 1/2000\}$. In order to simulate the dilute-system limit, the two vacancies must stay far apart, which justifies the choice of large $N_0$.  The distance between the two vacancies is thus monitored during the simulation.  The results are reported in Table~\ref{Tab:zeta_y_H} and shown in Fig.~\ref{Fig:zeta_y}. Note that for $N_0=2000$, the simulation box is no longer cubic but rectangular, leading to small anisotropies impacting the errors. The results confirm the linear dependence on the vacancy molar fraction $y$, even in the case of multiple vacancies in the dilute-system limit.

\subsubsection{Vacancy diffusion coefficient ${\cal D}_y$}

{Next, we compute the vacancy diffusion coefficient ${\cal D}_y$, which is the leading contribution to the diffusivity $D_{\rm vac}$ of the vacancy diffusion mode~(\ref{D_vac-final-1}). Using Eq.~\eqref {D-zeta} and the values of~$\zeta$ in Table~\ref{Tab:zeta_H} we obtain ${\cal D}_y$ as reported in Table~\ref{Tab:Dy_H} with the method of the Helfand moment.  Alternatively, ${\cal D}_y$ is obtained by the Markovian jump model for the random walk of a single vacancy in the crystal, as described in App.~\ref{app:Markov} and discussed in Sec.~\ref{ssec:llvd}. During the simulations performed for the computation of the transport coefficients, the jump times of the vacancy are recorded. Assuming that  the jump times are distributed according to an exponential probability distribution $p(\tau)=w\, {\rm e}^{-w\tau}$, we extract the vacancy jump rate $w$ and we obtain ${\cal D}_y$ from Eq.~\eqref{eq:D_omega}. The results are reported in Table~\ref{Tab:Dy_omega} and show an excellent agreement with the results obtained from the method of the Helfand moments, as seen in Fig.~\ref{Fig:D_y}.  On the one hand, Fig.~\ref{Fig:D_y}(a) shows that the vacancy diffusion coefficient quickly decreases as the crystal density $n_{0}$ increases, as a consequence of a similar behavior for the coefficient $\zeta$ in Fig.~\ref{Fig:TC-zeta}(a).  The results of Fig.~\ref{Fig:D_y}(a) about the dependence of ${\cal D}_y$ on the density $n_0$ are in good agreement with the literature \cite{BA71,vdMDF17}. On the other hand, in Fig.~\ref{Fig:D_y}(b), we observe that the vacancy diffusion coefficient converges to a positive value in the dilute-system limit $y\to 0$, which confirms the theoretical expectations.


\begin{figure*}[t!]\centering
{\includegraphics[width=.75\textwidth]{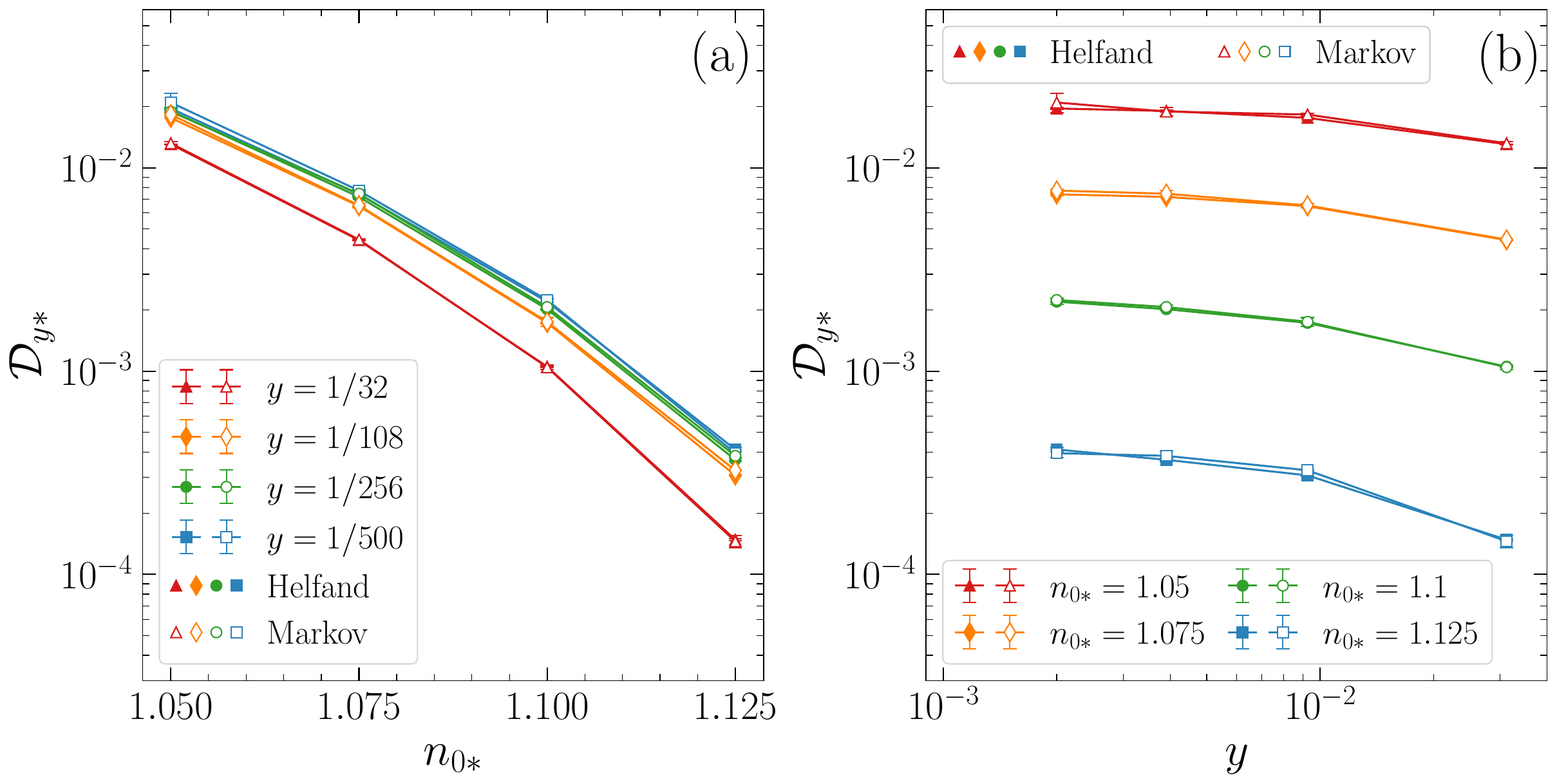}}
\caption[] {Vacancy diffusion coefficient ${\cal D}_y$ versus (a) the density $n_{0}$ and (b) the molar fraction of vacancy $y$. The  filled symbols correspond to ${\cal D}_y$ given in Table~\ref{Tab:Dy_H} and obtained from Eq.~\eqref{D-zeta}  and the method of the Helfand moments. The   open symbols correspond to ${\cal D}_y$ given in Table~\ref{Tab:Dy_omega}  and obtained from Eq.~\eqref{eq:D_omega}  and the Markovian jump model for the random walk of a single vacancy in the crystal.} \label{Fig:D_y}
\end{figure*}


\subsubsection{Single-vacancy spectral function}

In the case of a crystal with a single vacancy, we have numerically computed the time-dependent correlation functions~\eqref{eq:ISFV} with $N_{\rm v}=1$.  The periodic boundary conditions used in the simulation impose that the wave vector takes the discrete values ${\bf q} = 2\pi\left(n_x/L_x{\bf e}_x+n_y/L_y{\bf e}_y+n_z/L_z{\bf e}_z \right)$ with $(n_x,n_y,n_z)\in {\mathbb Z}^3$. For simplicity, we choose ${\bf q}$ in the $[100]$ direction with $n_x=1$, $n_y=n_z=0$. To perform the Fourier transform giving the spectral function~(\ref{eq:VSF-dfn}), it is necessary to obtain  the correlation function $F_{\rm vac}$ for a time interval that is long enough for the function to fully decay. The small values expected for the diffusion coefficients imply that these time intervals are typically large. To reduce the computational cost of the simulation, we consider only the cases $n_{0*}\in\{1.05,1.075\}$ and $y\in\{1/32,1/108,1/256\}$. The spectral function is obtained from a numerical Fourier transform of $F_{\rm vac}$.  The so-computed spectral function presents a resonance peak around zero frequency, which confirms the diffusive character of the vacancy diffusion mode.  The spectral function is fitted to a Lorentzian function of the form~(\ref{eq:VSF-Markov}) expected from the Markovian jump model.  For ${\bf q}$ in the $[100]$ direction, the half-width of the Lorentzian  is given by the dispersion relation~(\ref{Markov_disp_rel-fcc}) with ${\bf q}=(q,0,0)$, so that, in this case, the effective diffusivity takes the following value,
\be
\label{D_vac-eff-100}
D_{\rm vac}^{\rm (eff)}(q) = - \frac{1}{q^2}\, z(q,0,0) =  \frac{2w}{3 q^2} \bigg( 1 - \cos\frac{aq}{2} \bigg) .
\ee
Figure~\ref{Fig:spctrl_D_vac}(a) shows an example of the spectral function~(\ref{eq:VSF-dfn}) computed directly from molecular dynamics, together with two Lorentzian functions~(\ref{eq:VSF-Markov}). The first Lorentzian is computed with ${\cal D}_y$ from the Markovian jump model of Table~\ref{Tab:Dy_omega} as a parameter. We note that ${\cal D}_y$ is the zeroth-order contribution in $q$ to $D_{\rm vac}^{\rm (eff)}(q)$ given in~(\ref{D_vac-eff-100}). The second Lorentzian is computed with the exact theoretical value for $D_{\rm vac}^{\rm (eff)}(q)$  using the rates $w$ of the Markovian jump model. In Fig.~\ref{Fig:spctrl_D_vac}(b), the values of the effective vacancy diffusion coefficient obtained by fitting  Lorentzian functions to the computed spectral functions and given in Table~\ref{Tab:D_vac_SF} are plotted in comparison, first, of the leading-order diffusion coefficient given by Eq.~(\ref{eq:D_omega}) and, next, the expectation~(\ref{D_vac-eff-100}) from the exact dispersion relation of the Markovian jump process.  Since the latter includes corrections of orders higher than $O(q^2)$, the effective diffusion coefficient is smaller than its hydrodynamic limit for $q\to 0$, which is indeed observed in the numerical results plotted in Fig.~\ref{Fig:spctrl_D_vac}(b).  This is an indication that we are limited by the relatively small size $N_0$ of the system to a large value for the wave number~$q$, which is beyond the hydrodynamic regime.  Moreover, the large computational time of the simulations does not allow us to simulate more than a single vacancy.  Nevertheless, in the limit $y\to 0$ toward the largest possible systems in our simulations, the value of the positive diffusion coefficient ${\cal D}_y$ of our theory is indeed obtained.


\begin{figure*}[t!]\centering
{\includegraphics[width=.75\textwidth]{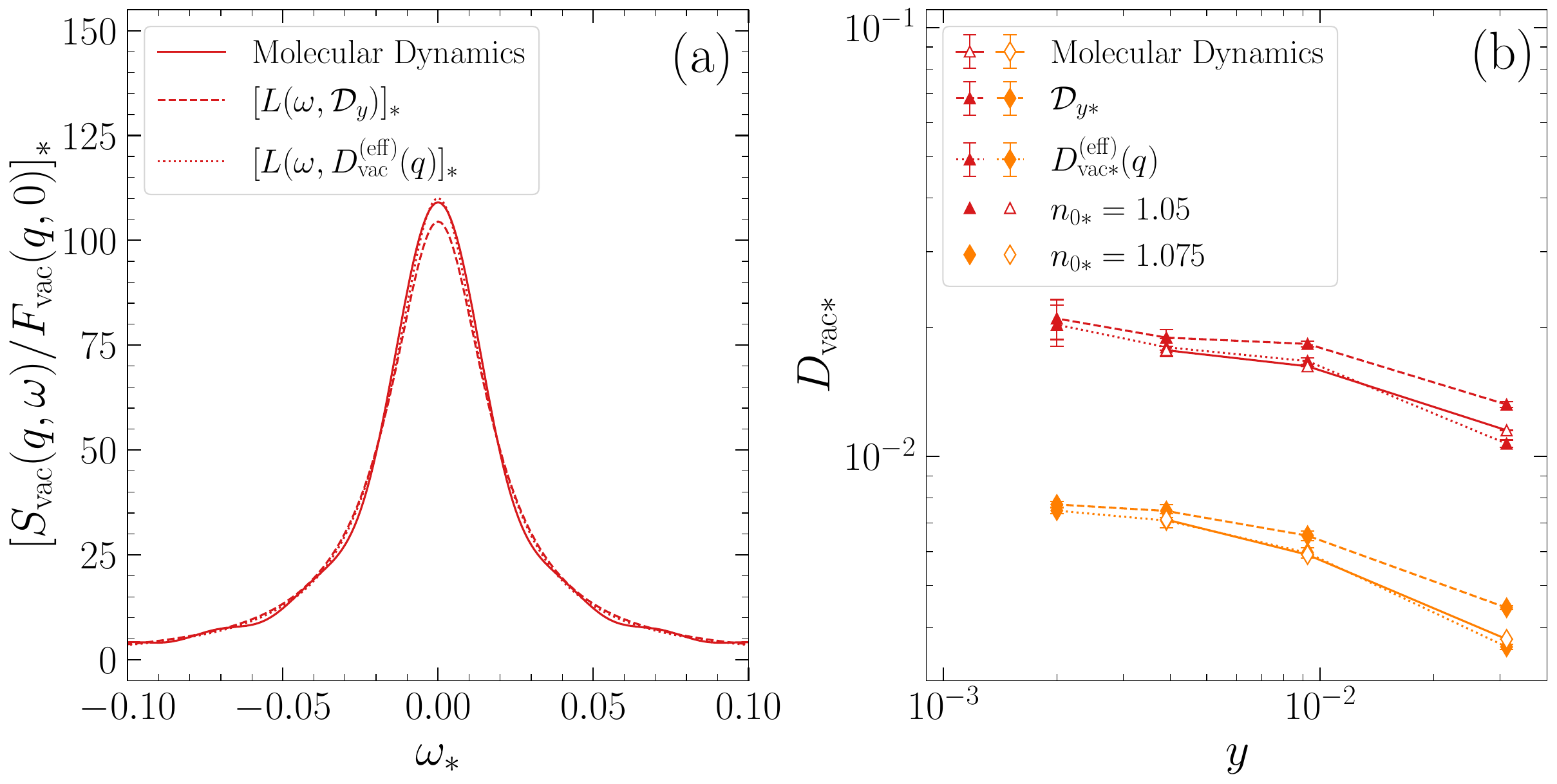}}
\caption[] { (a) Example of spectral function for the motion of a single vacancy in a crystal of density $n_{0*}=1.05$ and vacancy molar fraction $y=1/256$.  The spectral function is computed  by molecular dynamics for the wave vector taken in the direction $[100]$ with the magnitude $q_*=1.01$ (solid line).  The plot also shows the Lorentzian functions $L(\omega,D_{\rm vac})$ obtained with the value of ${\cal D}_y$ reported in Table~\ref{Tab:Dy_omega} (dashed line) and with the effective diffusivity $D_{\rm vac}^{\rm (eff)}(q)$ given by Eq.~(\ref{D_vac-eff-100}) (dotted line).
(b)~Diffusivity of the vacancy diffusion mode computed by molecular dynamics versus the molar fraction of vacancy $y$ for densities $n_{0*}\in\{1.05,1.075\}$ (solid lines).  These values given in Table~\ref{Tab:D_vac_SF} are obtained from the Lorentzian functions fitted to the computed spectral functions.  The values obtained from the fits are compared to the vacancy diffusion coefficient $D_{\rm vac}={\cal D}_y$ of the diffusivity at leading order given in Table~\ref{Tab:Dy_omega} (dashed lines) and to the expectation~(\ref{D_vac-eff-100}) from the Markovian jump model for the effective diffusivity $D_{\rm vac}^{\rm (eff)}(q)$ (dotted lines).} \label{Fig:spctrl_D_vac}
\end{figure*}


\subsubsection{Diffusivity $D_{\rm vac}$ of vacancy diffusion mode}

We then turn our interest to the diffusivity $D_{\rm vac}$ of the vacancy diffusion mode, which is predicted by hydrodynamics to take the value~(\ref{D_vac-final-1}) with the correction $\Delta D_{\rm vac}$ given by Eq.~(\ref{D_vac-final-2}).  We note that this prediction is for the hydrodynamic regime since $D_{\rm vac}$ is the coefficient of the term of $O(q^2)$ in the dispersion relation. This coefficient depends on the direction of the wave vector, but not on its magnitude.  The dependence on the direction ${\bf e}_{\rm l}={\bf q}/q$ is due to the dependence of $\Delta D_{\rm vac}$ on the  coefficient $B_T^{\rm l}$ and, thus, on the longitudinal speed of sound $c_{\rm l}$, which is known to depend on the direction of the wave vector.

In order to evaluate the correction $\Delta D_{\rm vac}$, we need the thermodynamic and elastic properties entering the  formula~(\ref{D_vac-final-2}).  The values of these properties are reported in Table~\ref{Tab:TH_EC}. The pressure is calculated using the method described in Ref.~\cite{MG24a}. For each density, the pressure is obtained for fixed $y$ and increasing values of $N_0$. An extrapolation to infinite $N_0$ gives the pressure as function of $y$. A second extrapolation gives the pressure at $y=0$. These values are in agreement with the ones obtained for the perfect crystal~\cite{MG24a,S98}. The extrapolation to $y=0$  is obtained with a linear least square regression, the slope of the interpolated curve provides the values of $\pi_y$. The equation of state for the pressure gives the isothermal bulk modulus $B_{T}$ defined by Eq.~(\ref{B_T}) and the specific heat capacity $c_{p}$ at constant pressure~\cite{MG24b}. The specific heat ratio  $\gamma$ is obtained from $c_{p}$ and using that $c_{v}=3k_{\rm B}/(2m)$ for hard spheres.  The product $\alpha B_T$ is calculated from Eq.~(\ref{dsigma/dT}).  The elastic constants are obtained with the method described in Ref.~\cite{MG24a}. The precision attained by our simulation is not sufficient to observe any significant deviation between the elastic  constants computed for a perfect crystal and a crystal with vacancies. The vacancy concentration $y_{\rm eq}$ is obtained from the data published in Ref.~\cite{L20}. Finally, the derivative $\varsigma_y$ of the specific entropy  with respect to the vacancy molar fraction is calculated using Eq.~\eqref{dSdy/Na-HS-2} and the data of Table~\ref{Tab:TH_EC}.


\begin{figure}[t!]\centering
{\includegraphics[width=.4\textwidth]{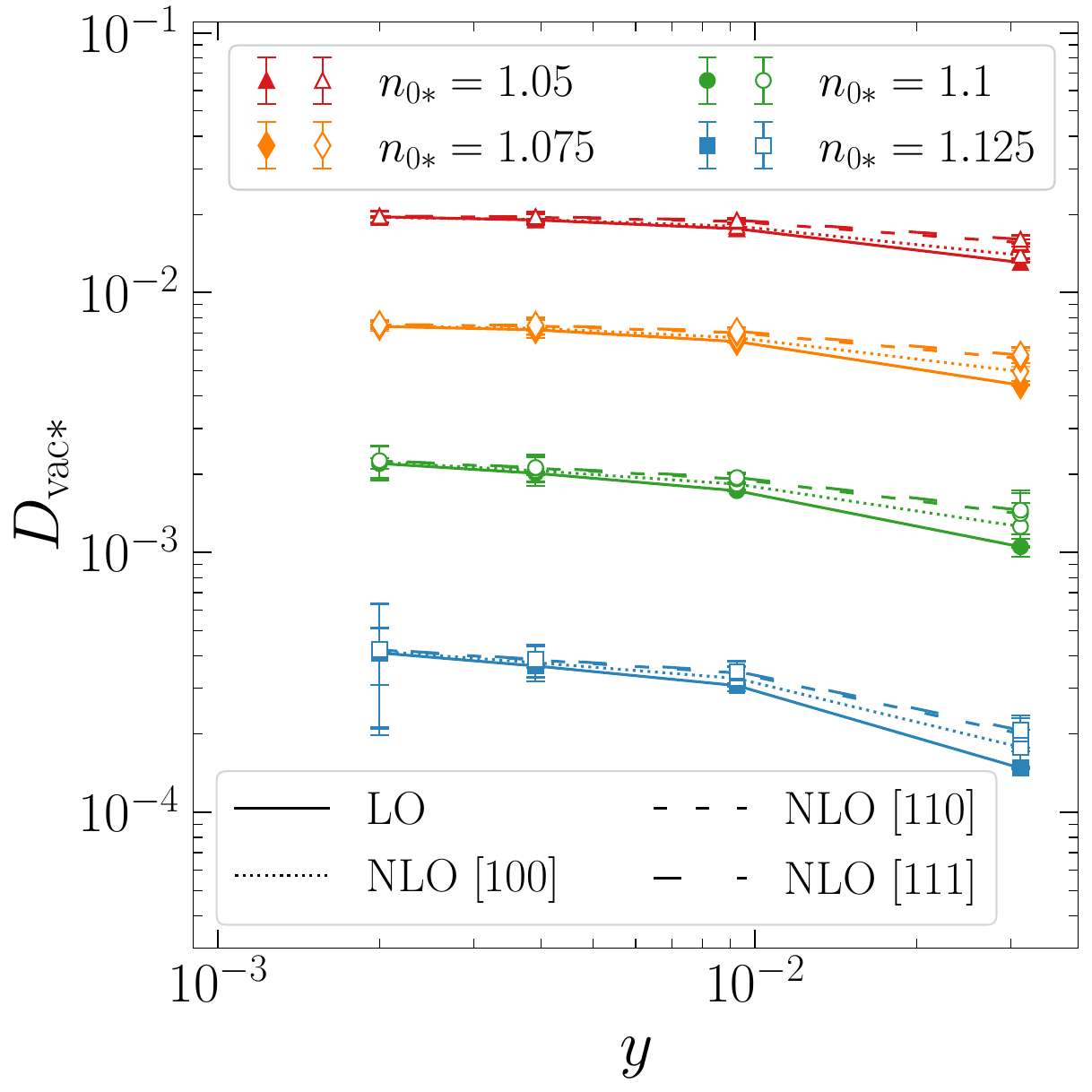}}
\caption[] {Diffusivity ${D}_{\rm vac}$ of the vacancy diffusion mode versus the molar fraction of vacancy $y$ for densities $n_{0*}\in\{1.05,1.075\}$. $D_{\rm vac}={\cal D}_y$ at leading order (LO, filled symbols) is given in Table~\ref{Tab:Dy_H}. $D_{\rm vac}={\cal D}_y+\Delta D_{\rm vac}$ at next-to-leading order (NLO, open symbols) is obtained in the three directions $[100]$, $[110]$, and $[111]$ of the wave vector $\bf q$ (dotted, dashed, and long dashed lines) by using Eqs.~\eqref{D_vac-final-1}-\eqref{D_vac-final-2} with Eq.~\eqref{dSdy/Na-HS-2} for the derivative $\varsigma_y$ of the specific entropy with respect to the vacancy molar fraction, and the data of Table~\ref{Tab:TH_EC}.} \label{Fig:D_vac}
\end{figure}

The next-to-leading-order contribution to $D_{\rm vac}$ given by Eq.~(\ref{D_vac-final-2})
is calculated  using the transport coefficients and the thermodynamic and elastic properties given in Tables~\ref{Tab:zeta_H},~\ref{Tab:xi_H} and~\ref{Tab:TH_EC}. The results are shown in Fig.~\ref{Fig:D_vac} for the wave vector $\bf q$ in the three special directions $[100]$, $[110]$, and $[111]$, in which the hydrodynamic equations can be decoupled. First of all, the data are in agreement with the expected $y$-behavior of $\Delta D_{\rm vac}$, which should scale as $O(y\ln y)$ and vanish as $y\to 0$.  We note that the error on the numerical results increases significantly with $n_{0}$ and $y^{-1}$. These errors are a direct consequence of the large numerical errors on $\xi$ discussed previously.  Furthermore, we note that  the diffusivity of the vacancy mode has a relatively small dependence on the direction of the wave vector.

In principle, the diffusivity $D_{\rm vac}$ of the vacancy diffusion mode could be numerically computed from the correlation and spectral functions associated with the microscopic density of vacancies $\hat{y}$, as defined in App.~\ref{app:S_vac}.  These functions have their poles at complex frequencies, which depend on the direction of the wave vector $\bf q$.  Since the positions of the vacancies are determined by the microscopic dynamics of the particles, the vacancy concentration is expected to be ruled on large spatiotemporal scales by the eight coupled hydrodynamic equations, thus also involving the other macrofields.  However, in order to reach the hydrodynamic limit, the number of vacancies should be increased as $N_{\rm v}=N_0 y \gg 1$ with the system size $N_0\gg 1$, while keeping positive and constant the vacancy molar fraction $y=N_{\rm v}/N_0$.  This requires more important computational resources than available to us and the computation of vacancy spectral functions probing the collective motion of several interacting vacancies should be left for future work.

In any case, we observe that the results tends to converge toward a finite value in the limit of small $y$, providing evidence for a non-vanishing diffusion coefficient of vacancies in solids.


\section{Conclusion and perspectives}
\label{sec:conclusion}

In this work, we have developed the hydrodynamics of nonperfect monatomic crystals, clarifying the relationship between the known transport coefficients of the crystal and the vacancy diffusion coefficient.  The approach is based on the chemical potential of vacancies, which should be introduced within the local-equilibrium thermodynamics of nonperfect monatomic crystals to determine the thermodynamic forces or affinities driving the crystal out of equilibrium.  Using the constraint between the concentration of vacancies, the strain tensor, and the density of mass, the Gibbs and Euler thermodynamic relations for the energy density can be transformed to eliminate the term associated with the vacancies.  Consequently, the chemical potential of atoms, the excess of the stress tensor, and the pressure are modified with a contribution from the chemical potential of vacancies.  With this modification, the hydrodynamics of crystals with vacancies finds a formulation similar as for perfect crystals.  The dissipative current densities contributing to entropy production can be deduced and expressed in terms of the modified thermodynamic forces or affinities.  The approach is justified by the local-equilibrium approach to statistical mechanics starting from the underlying microscopic dynamics of the atoms composing the crystal.  In this way, Green-Kubo and Einstein-Helfand formulas are obtained for the transport coefficients of crystals with vacancies.  

In addition to the rank-four tensor of viscosities, there are the rank-two tensors of the heat conductivities, the coefficients of vacancy thermodiffusion, and the vacancy conductivities.  The latter are the coefficients coupling the dissipative current density of vacancies to the gradient of the excess of stress tensor driving the flow of vacancies in a stressed crystal.  In this regard, the vacancy conductivities should vanish with the concentration of vacancies.  However, the modification of the excess of stress tensor by the chemical potential of vacancies has the consequence of adding a Fickian term of genuine vacancy diffusion to the current density of vacancies.  The so-defined Fickian vacancy diffusion coefficient~(\ref{D_y}) is proportional to the vacancy conductivity by a factor equal to the derivative of the chemical potential of vacancies with respect to their concentration (or molar fraction).  Since this chemical potential goes as the logarithm of the vacancy concentration in the limit of a dilute system of vacancies, the factor has the right dependence on the vacancy concentration to compensate the one of the vacancy conductivity and to find a vacancy diffusion coefficient remaining finite and positive in the dilute-system limit.  Therefore, the vacancy conductivity is directly related to the vacancy diffusion coefficient, which is now precisely identified within the framework of the hydrodynamics of crystals.

Next, the coupled equations of crystal hydrodynamics have been obtained and linearized around equilibrium to deduce the eight hydrodynamic modes and their dispersion relation for crystals with vacancies.  The seven modes of sound propagation and heat conduction have the dispersion relations of the perfect crystal with a correction negligible in the limit of an arbitrarily low vacancy concentration.  The eighth mode of vacancy diffusion now has a nontrivial diffusivity given by the positive and finite Fickian coefficient of vacancy diffusion as leading contribution and a next-to-leading contribution vanishing with the vacancy concentration.  In the limit of low vacancy concentration, the interaction between the vacancies disappears and their motion can be modeled with a Markovian jump stochastic process giving the expected leading contribution to the vacancy diffusion coefficient.

The theoretical predictions are confirmed by numerical simulations of the hard-sphere crystal with vacancies.  We compute the vacancy conductivity, the coefficient of vacancy thermodiffusion, and the vacancy diffusion coefficient as functions of the density of lattice sites and the vacancy molar fraction.  The numerical results confirm the proportionality of the vacancy conductivity and the coefficient of vacancy thermodiffusion with respect to the vacancy molar fraction in the limit of zero vacancy molar fraction, while the vacancy diffusion coefficient keeps a positive and finite value in this limit.  Furthermore, the vacancy spectral function is computed to evaluate the diffusivity for the motion of a single vacancy, showing agreement with theoretical predictions.  Finally, using the numerically computed values of the thermodynamic quantities and transport coefficients, the hydrodynamic dispersion relation of vacancies is calculated in the three main lattice directions of the wave vector. Again, the diffusivity converges toward the leading contribution given by the vacancy diffusion coefficient in the limit of vanishing vacancy molar fraction.  The results show a small anisotropy of the diffusivity in the different lattice directions.

In conclusion, the numerical results support the theoretical predictions of the hydrodynamics of nonperfect monatomic crystals.  In particular, we have obtained numerical evidence for the phenomenon of vacancy thermodiffusion in the hard-sphere nonperfect crystal.  The relationship between the previously known coefficients of vacancy transport and the Fickian coefficient of vacancy diffusion has been clarified.

Our work opens important perspectives in the study of the transport of vacancies and other point defects in crystals by {\it ab initio} methods based on the microscopic dynamics of the atoms composing the crystal.  The theory shows how the chemical potential of point defects should modify the excess of stress tensor and the pressure in order to understand how the presence of  point defects can be formulated in the thermodynamics and the elastodynamics of nonperfect crystals.  The methods we have here developed can be extended {\it mutatis mutandis} from classical to quantum mechanics \cite{MG23}.


\section*{Acknowledgements}

The authors thank James F. Lutsko for fruitful discussions and for communicating us the data published in Ref.~\cite{L20} on vacancy concentration.  The authors acknowledge the support of the Universit\'e Libre de Bruxelles (ULB) and the Fonds de la Recherche Scientifique de Belgique (F.R.S.-FNRS) in this research. J.~M. is a Postdoctoral Researcher of the Fonds de la Recherche Scientifique de Belgique (F.R.S.-FNRS).  Computational resources have been provided by the Consortium des Equipements de Calcul Intensif (CECI), funded by the Fonds de la Recherche Scientifique de Belgique (F.R.S.-FNRS) under Grant No. 2.5020.11 and by the Walloon Region.

\appendix

\section{Equilibrium thermodynamics of a crystal with vacancies}
\label{app:thermo}

\subsection{Generalities}

We consider a crystal in equilibrium at the temperature $T$, in the volume $V$, with $N$ atoms and $N_{\rm v}$ vacancies.  The total number of lattice sites is thus equal to $N_0=N+N_{\rm v}$.  The volume of a lattice site is defined as $v\equiv V/N_0$, which is given by $v=1/n_{{\rm eq},0}$ in terms of the equilibrium density of lattice sites,  $n_{{\rm eq},0}=n_0$.  At equilibrium, the molar fraction of vacancies takes the value $y=N_{\rm v}/N_0$.  The Helmholtz free energy $F\equiv E-TS$ of this nonperfect crystal obeys the following Gibbs and Euler relations,
\begin{align}
dF &= - S \, dT - p \, dV + \mu_{\rm a}\, dN + \mu_{\rm v}\, dN_{\rm v} \, , \label{F-Gibbs}\\
F &= - p \, V + \mu_{\rm a}\, N + \mu_{\rm v}\, N_{\rm v} \, , \label{F-Euler}
\end{align}
where $E$ is the energy, $S$ is the entropy, $p$ the pressure, $\mu_{\rm a}$ the atomic chemical potential, and $\mu_{\rm v}$ the vacancy chemical potential.  These chemical potentials have been introduced in Sec.~\ref{sec:thermo}. 

These thermodynamic quantities can be evaluated using the equilibrium statistical mechanics of crystals with vacancies \cite{HL64,AL87,LC85,PF01}.

\subsection{The crystal at low vacancy density}

If the density of vacancies is low, i.e., if $1\ll N_{\rm v}\ll N$ or $y\ll 1$, the configurational entropy of the vacancies can be evaluated with Boltzmann's formula as
\be\label{DS_conf}
\Delta S = k_{\rm B}\ln \frac{N_0!}{N! \, N_{\rm v}!} \simeq - N_0 k_{\rm B} \left[ (1-y)\ln(1-y)+ y \ln y\right] .
\ee
Moreover, we may introduce the free energy $f_0(T,v)$ per atom in the perfect crystal and the free energy of formation of a vacancy in the crystal as $\tilde f_{\rm v}(T,v,y) = f_{\rm v}(T,v)+O(y)$, where the correction of order $y$ is due to the mutual interaction between the vacancies, $f_{\rm v}(T,v)$ being the free energy of formation of a vacancy in the perfect crystal.  Accordingly, the total free energy is given by
\begin{align}
N_0^{-1} F = &\ (1-y) \, f_0(T,v) + y \, \tilde f_{\rm v}(T,v,y) \nonumber\\
& + k_{\rm B}T \left[ (1-y)\ln(1-y)+ y \ln y\right] .
\end{align}
We thus   infer that the pressure and its derivative with respect to the vacancy molar fraction can be expressed as
\begin{align}
p &= - \left(\frac{\partial f_0}{\partial v}\right)_T + O(y) \, , \\
\left(\frac{\partial p}{\partial y}\right)_{T,v} &= \left(\frac{\partial f_0}{\partial v}\right)_T  - \left(\frac{\partial f_{\rm v}}{\partial v}\right)_T + O(y) \, ,
\end{align}
the atomic chemical potential as
\be
\mu_{\rm a} = f_0 - v \left(\frac{\partial f_0}{\partial v}\right)_T + O(y) \, ,
\ee
and the vacancy chemical potential as
\be
\label{mu_v}
\mu_{\rm v} = g_{\rm v} + k_{\rm B} T \, \ln y + O(y) \, ,
\ee
where
\be
\label{g_v}
g_{\rm v} = f_{\rm v} + v \, p(T,v,0) = f_{\rm v}  - v \left(\frac{\partial f_0}{\partial v}\right)_T
\ee
is the Gibbs free energy of formation of a vacancy.  If the number of vacancies is not conserved, for instance, if they can move to the surface of the crystal, the vacancy chemical potential should be equal to zero at equilibrium, $\mu_{\rm v,eq} = 0$, which shows that the equilibrium value of the vacancy molar fraction can be evaluated as
\be
\label{y_eq}
y_{\rm eq}(T,v) \simeq {\rm e}^{-\beta g_{\rm v}} = {\rm e}^{-\beta(f_{\rm v}+vp)}
\ee
with $\beta=(k_{\rm B}T)^{-1}$.  Therefore, the knowledge of the equilibrium vacancy molar fraction leads to the knowledge of their Gibbs free energy of formation.

Furthermore, we also obtain the following expressions for the entropy per atom and its derivative with respect to the vacancy molar fraction,
\begin{align}
\frac{S}{N} =& - \left(\frac{\partial f_0}{\partial T}\right)_v - y \left(\frac{\partial f_{\rm v}}{\partial T}\right)_v+ k_{\rm B} (y-y\ln y) \nonumber\\
& + O(y^2\ln y) \, , \label{S/Na} \\
\frac{\partial}{\partial y}\left(\frac{S}{N}\right)_{T,v} =& - \left(\frac{\partial f_{\rm v}}{\partial T}\right)_v - k_{\rm B} \ln y + O(y\ln y) \, . \label{dSdy/Na}
\end{align}

\subsection{The hard-sphere nonperfect crystal}

In hard-sphere systems, the total energy is given by $E=\frac{3}{2} N k_{\rm B}T$ and the pressure has the form $p=k_{\rm B}T \varphi(v,y)$ with some function $\varphi(v,y)$ that can be obtained from numerical calculations.  Moreover, the equilibrium value of the vacancy molar fraction does not depend on the temperature and can also be obtained using numerical calculations, giving $y_{\rm eq}(v)$ \cite{L20}.  Therefore, in hard-sphere crystals, the vacancy chemical potential takes the following form,
\be\label{m_v-HS}
\mu_{\rm v} = k_{\rm B}T \, \ln \frac{y}{y_{\rm eq}(v)} + O(y) \, .
\ee
Using Eq.~(\ref{y_eq}), the Helmholtz free energy of vacancy formation is given by
\be
f_{\rm v} = -k_{\rm B}T \ln y_{\rm eq}(v) - v \, k_{\rm B}T \, \varphi(v,0) + O(y_{\rm eq}) \, .
\ee
Substituting this result into Eq.~(\ref{dSdy/Na}), we thus find 
\begin{align}
\label{dSdy/Na-HS}
\frac{\partial}{\partial y}\left(\frac{S}{N}\right)_{T,v} =&\  k_{\rm B}\, v \,  \varphi(v,0) - k_{\rm B} \ln\frac{y}{y_{\rm eq}(v)} \nonumber\\
& + O(y_{\rm eq}) + O(y\ln y) \, ,
\end{align}
for the hard-sphere nonperfect crystal.  Equivalently, the derivative of the specific entropy with respect to the vacancy molar fraction is obtained as
\be
\label{dSdy/Na-HS-2}
\varsigma_y=\left(\frac{\partial\mathfrak{s}}{\partial y}\right)_{T,v} = \frac{1}{m T} \left[ v \, p(T,v,0) - \mu_{\rm v}(T,v,y)\right] + O(y\ln y)
\ee
with $p(T,v,0)=k_{\rm B}T\, \varphi(v,0)$ and $\mu_{\rm v}$ given by Eq.~(\ref{m_v-HS}).


\section{Vacancy Helfand moment ${\mathbb G}^a_y$} 
\label{app:Gy}

The microscopic molar fraction of vacancies  being defined by Eq.~(\ref{y-micro}), the microscopic current density associated  with $\hat{y}$ is obtained from
\begin{align}
\partial_t\,\hat{y}({\bf r},t)& =-\frac{1}{n_{\rm eq, 0}}\sum_{i=1}^{N_{\rm v}}\frac{d{\bf r}_{{\rm v},i}(t)}{dt}\cdot{\boldsymbol \nabla}\, \delta\big[{\bf r}-{\bf r}_{{\rm v},i}(t)\big]\notag\\
&=-{\boldsymbol \nabla}\cdot\bigg\{\frac{1}{n_{\rm eq, 0}}\sum_{i=1}^{N_{\rm v}} {\bf v}_{{\rm v},i}(t)\, \delta\big[{\bf r}-{\bf r}_{{\rm v},i}(t)\big]\bigg\}\notag\\
&=-{\boldsymbol \nabla}\cdot{\hat{\bf J}}_y({\bf r},t) \, ,
\end{align}
where ${\bf v}_{{\rm v},i}\equiv d{\bf r}_{{\rm v},i}/dt$ is the velocity of the vacancy $i$. Using that $\langle \hat{ \bf J}_y \rangle_{{\rm eq}}=0$ in the frame moving with the element of the medium, the Helfand moment  for the motion of vacancies given in Eq.~\eqref{Helfand} becomes
\begin{align}
{\mathbb G}^a_y(t)&= \int_0^t d\tau\int_Vd{\bf r}\ \hat{J}^{a}_y({\bf r},\tau)\notag\\
&=\int_0^t d\tau\int_Vd{\bf r}\ \frac{1}{n_{\rm eq, 0}}\sum_{i=1}^{N_{\rm v}} {v}_{{\rm v},i}^{a}(\tau)\, \delta\big[{\bf r}-{\bf r}_{{\rm v},i}(\tau)\big]\notag\\
&=\frac{1}{n_{\rm eq, 0}}\int_0^t d\tau\ \sum_{i=1}^{N_{\rm v}} \frac{d{r}_{{\rm v},i}^{a}(\tau)}{d\tau}\notag\\
&=\frac{1}{n_{\rm eq, 0}}\sum_{i=1}^{N_{\rm v}} \left[ {r}_{{\rm v},i}^{a}(t)-{r}_{{\rm v},i}^{a}(0)\right] .
\label{Helfand-y}
\end{align}
Since the position of the vacancy is always on a lattice site, the distance  $\Vert{\bf r}_{{\rm v},i}(t)-{\bf r}_{{\rm v},i}(0)\Vert$ travelled by a vacancy is taken with respect to the lattice, which itself moves due to the conservation of the center of mass position as explained in Sec.~\ref{ssec:NPHSC}. We thus interpret $\Vert{\bf r}_{{\rm v},i}(t)-{\bf r}_{{\rm v},i}(0)\Vert$ as the distance on the lattice separating the sites, where the vacancy is located at times $t_0=0$ and $t$.

We note that, in the case of a single vacancy, the standard Einstein formula for the vacancy diffusion coefficient in terms of the mean square displacement of the vacancy~\cite{E26} is recovered by substituting the Helfand moment~(\ref{Helfand-y}) with $N_{\rm v}=1$ into the Einstein-Helfand formula~(\ref{zeta-H-y}) and using the definition~(\ref{D_y}).


\section{Calculation of the hydrodynamic modes}
\label{app:hydro-modes}

The closed set of linear equations (\ref{macro-eq-y-fin})-(\ref{macro-eq-u-fin}) are solved by spatial Fourier transforms with the method we developed in Ref.~\cite{MG21}, but here taking $\nabla^a=-{\rm i}\, q^a$ as in Ref.~\cite{MG24b}.  Furthermore, the equations can be projected onto the directions of the orthonormal basis $\{ {\bf e}_{\rm l},{\bf e}_{{\rm t}_1},{\bf e}_{{\rm t}_2}\}$ consisting of the unit vectors such that ${\bf e}_{\rm l}\equiv{\bf q}/\Vert{\bf q}\Vert$ and ${\bf e}_{{\rm t}_k}\cdot{\bf q}=0$ for $k=1,2$, and leading to the expansions $\delta v^a = \sum_\sigma \delta v^\sigma e^a_\sigma$ and  $\delta u^a = \sum_\sigma \delta u^\sigma e^a_\sigma$.

In this way, the eight hydrodynamic equations are written as
\be\label{eq-psi}
\partial_t \, \pmb{\psi} = {\boldsymbol{\mathsf L}}\cdot\pmb{\psi}  \quad \mbox{with} \quad
\pmb{\psi}=(\delta{y},\delta{\mathfrak s},\delta v^\sigma,\delta u^\sigma)^{\rm T}
\ee
in terms of the $8\times 8$ matrix
\be
{\boldsymbol{\mathsf L}} = {\boldsymbol{\mathsf L}}^{(0)} + {\boldsymbol{\mathsf L}}^{(1)}
\ee
decomposed into its adiabatic part
\be
{\boldsymbol{\mathsf L}}^{(0)}
=
\left[
\begin{array}{cccc}
0 & 0 & 0 & 0 \\
0 & 0 & 0 & 0 \\
-{\rm i} C_{vy} \delta^{{\rm l}\sigma} q & -{\rm i} C_{v{\mathfrak s}} \delta^{{\rm l}\sigma} q & 0& -A^{\sigma\sigma'}  q^2 \\
0 & 0 & \delta^{\sigma\sigma'} & 0 \\
\end{array}
\right]
\ee
and its dissipative part
\be
{\boldsymbol{\mathsf L}}^{(1)}
=
\left[
\begin{array}{cccc}
-D_{yy} q^2 & -D_{y{\mathfrak s}} q^2 & 0 & {\rm i} D_{yu}^{{\rm l}\sigma'} q^3 \\
-D_{{\mathfrak s}y} q^2 & -D_{\mathfrak{ss}} q^2 & 0 & {\rm i} D_{{\mathfrak s}u}^{{\rm l}\sigma'} q^3 \\
0 & 0 & -D_{vv}^{\sigma\sigma'} q^2 & 0 \\
-{\rm i} D_{uy} \delta^{{\rm l}\sigma} q & -{\rm i} D_{u{\mathfrak s}} \delta^{{\rm l}\sigma} q & 0 & -D_{uu}^{\sigma\sigma'} q^2 \\
\end{array}
\right]
\ee
with the following coefficients,
\begin{align}
& A^{\sigma\sigma'} \equiv C_{vu}^{abcd} e_{\rm l}^a e^b_\sigma e_{\rm l}^c e^d_{\sigma'} \, , \nonumber\\
& D_{yu}^{\sigma\sigma'} \equiv D_{yu}^{abcd} e_{\rm l}^a e^b_\sigma e_{\rm l}^c e^d_{\sigma'} \, , \quad
D_{{\mathfrak s}u}^{\sigma\sigma'} \equiv D_{{\mathfrak s}u}^{abcd} e_{\rm l}^a e^b_\sigma e_{\rm l}^c e^d_{\sigma'} \, , \nonumber\\
& D_{vv}^{\sigma\sigma'} \equiv D_{vv}^{abcd} e_{\rm l}^a e^b_\sigma e_{\rm l}^c e^d_{\sigma'} \, , \quad
D_{uu}^{\sigma\sigma'} \equiv D_{uu}^{abcd} e_{\rm l}^a e^b_\sigma e_{\rm l}^c e^d_{\sigma'} \, .
\label{A2-D2}
\end{align}
As in Ref.~\cite{MG24b}, we may consider the special directions $[100]$, $[110]$, and $[111]$, where the $3\times 3$ matrices $\big[ A^{\sigma\sigma'}\big]$ and $\big[D_{vv}^{\sigma\sigma'}\big]$ commute.  In these directions, they can be simultaneously diagonalized, giving
\be\label{diagonal-A-Dvv}
A^{\sigma\sigma'} = A^{\sigma} \, \delta^{\sigma\sigma'}
\quad\mbox{and}\quad
D_{vv}^{\sigma\sigma'} = D_{vv}^{\sigma} \, \delta^{\sigma\sigma'}  .
\ee

The eigenvalue problem ${\boldsymbol{\mathsf L}}\cdot\pmb{\psi} = z\, \pmb{\psi}$ can be solved by expanding the eigenvalues and the eigenvectors as $z= z^{(0)} + z^{(1)} + \cdots$ and $\pmb{\psi} = \pmb{\psi}^{(0)} +\pmb{\psi}^{(1)} + \cdots$.  We first solve the eigenvalue problem at zeroth order to get the eight eigenvalues and the corresponding right and left eigenvectors:
\be
{\boldsymbol{\mathsf L}}^{(0)}\cdot\pmb{\psi}^{(0)} = z^{(0)}\, \pmb{\psi}^{(0)} \, , \quad
{\pmb{\tilde\psi}}^{(0)\dagger} \cdot {\boldsymbol{\mathsf L}}^{(0)}= z^{(0)}\, {\pmb{\tilde\psi}}^{(0)\dagger}  \, .
\ee
The left and right eigenvectors are taken to satisfy the biorthonormality conditions, $\pmb{\tilde\psi}^{(0)\dagger}_{\alpha}\cdot\pmb{\psi}^{(0)}_{\beta}=\delta_{\alpha\beta}$ for $\alpha,\beta=1,2,\dots,8$.  
They are given by
\begin{align}
& z^{(0)}_{\alpha} = \pm{\rm i} \, \sqrt{A^\sigma} q \, , \quad \
\pmb{\psi}^{(0)}_{\alpha} =
\left(
\begin{array}{c}
0 \\
0 \\
\pm{\rm i} \, \sqrt{A^\sigma} q\, {\bf 1}_{\sigma} \\
{\bf 1}_{\sigma} 
\end{array}
\right) , \nonumber\\
& 
\pmb{\tilde\psi}^{(0)}_{\alpha} =\frac{1}{2}
\left(
\begin{array}{c}
-{\rm i} \, \frac{C_{vy}}{A^\sigma q}\, \delta^{{\rm l}\sigma} \\
-{\rm i} \, \frac{C_{v{\mathfrak s}}}{A^\sigma q}\, \delta^{{\rm l}\sigma} \\
\pm{\rm i} \, \frac{1}{\sqrt{A^\sigma} q}\, {\bf 1}_{\sigma} \\
{\bf 1}_{\sigma} 
\end{array}
\right) \quad\mbox{for} \ \ \alpha=1,2,\dots,6\, ; 
\label{modes-1-6}
\\
& z^{(0)}_{7} = 0 \, , \ \ 
\pmb{\psi}^{(0)}_{7} =
\left(
\begin{array}{c}
0 \\
1 \\
{\bf 0} \\
-{\rm i}\, \frac{C_{v{\mathfrak s}}}{A^{\rm l} q} \, {\bf 1}_{\rm l} 
\end{array}
\right) , \ \ 
\pmb{\tilde\psi}^{(0)}_{7} =
\left(
\begin{array}{c}
0 \\
1 \\
{\bf 0} \\
{\bf 0}\end{array}
\right) ; 
\label{z0_7}\\
& z^{(0)}_{8} = 0 \, , \ \ 
\pmb{\psi}^{(0)}_{8} =
\left(
\begin{array}{c}
1 \\
0 \\
{\bf 0} \\
-{\rm i}\, \frac{C_{vy}}{A^{\rm l} q} \, {\bf 1}_{\rm l}
\end{array}
\right) , \ \ 
\pmb{\tilde\psi}^{(0)}_{8} =
\left(
\begin{array}{c}
1 \\
0 \\
{\bf 0} \\
{\bf 0}\end{array}
\right) ;
\qquad\ \  \label{z0_8}
\end{align}
where $({\bf 1}_{\sigma})^{\sigma'}\equiv \delta^{\sigma\sigma'}$.  In Eq.~(\ref{modes-1-6}), we have the plus sign with $\sigma={\rm l},{\rm t}_1,{\rm t}_2$ for $\alpha=1,2,3$; and the minus sign again with $\sigma={\rm l},{\rm t}_1,{\rm t}_2$ for $\alpha=4,5,6$.  Accordingly, the modes $\alpha=1$-$6$ are the propagative sound modes with $z_{\alpha}^{(0)}=O(q)$, $\alpha=7$ is the heat conduction mode, and $\alpha=8$ the vacancy diffusion mode.  Using Eqs.~(\ref{C_vu}) and~(\ref{B_s}), the speeds of the sound modes take their expected values because $A^\sigma$ are the eigenvalues of the matrix
\be
A^{\sigma\sigma'} = \frac{1}{\rho} \, B_{\mathfrak s}^{abcd} e^a_{\rm l} e^b_\sigma e^c_{\rm l} e^d_{\sigma'} = \frac{1}{\rho} \big[B_{T}^{\sigma\sigma'} + (\gamma-1) B_T \, \delta^{{\rm l}\sigma}\delta^{{\rm l}\sigma'} \big]
\ee
with $B_{T}^{\sigma\sigma'}\equiv B_{T}^{abcd} e_{\rm l}^a e^b_\sigma e_{\rm l}^c e^d_{\sigma'}$, so that its eigenvalues are given by
\be
A^{\rm l} = \frac{1}{\rho} \left[B_{T}^{\rm l} + (\gamma-1) B_T\right] = c_{\rm l}^2
\label{A-l}
\ee
and
\be
A^{{\rm t}_k} = \frac{1}{\rho}\, B_{T}^{{\rm t}_k} = c_{{\rm t}_k}^2  
\qquad\mbox{with}\quad k=1,2 \, ,
\label{A-t}
\ee
since $B_{T}^{\sigma\sigma'}=B_{T}^{\sigma}\,\delta^{\sigma\sigma'}$ in the special directions where Eq.~(\ref{diagonal-A-Dvv}) holds.

Next, we calculate the first-order corrections giving the dissipative contributions to the dispersion relations, using
\be
z^{(1)}_{\alpha} = \frac{\pmb{\tilde\psi}^{(0)\dagger}_{\alpha}\cdot{\boldsymbol{\mathsf L}}^{(1)}\cdot\pmb{\psi}^{(0)}_{\alpha}}{\pmb{\tilde\psi}^{(0)\dagger}_{\alpha}\cdot\pmb{\psi}^{(0)}_{\alpha}} \, .
\ee
We obtain
\begin{align}
z^{(1)}_{\alpha} &= z^{(1)}_{\alpha+3}  \nonumber\\
&=  -\frac{1}{2} \left(D_{vv}^{\sigma} + D_{uu}^{\sigma\sigma} + \frac{ C_{vy} D_{yu}^{\rm ll} + C_{v{\mathfrak s}} D_{{\mathfrak s}u}^{\rm ll} }{A^{\rm l}}\, \delta^{{\rm l}\sigma} \right) q^2 \nonumber\\
& \qquad\qquad\qquad\quad\mbox{for}\quad \alpha=1,2,3 \, ,  \label{z1_123+456}\\
z^{(1)}_{7} &= - \left( D_{\mathfrak{ss}} -  \frac{C_{v{\mathfrak s}}}{A^{\rm l}}\, D_{{\mathfrak s}u}^{\rm ll} \right) q^2  \, , \label{z1_7}\\
z^{(1)}_{8} &= - \left( D_{yy} -  \frac{C_{vy}}{A^{\rm l}}\, D_{yu}^{\rm ll} \right) q^2  \, , \label{z1_8}
\end{align}
where there is no summation over repeated indices.  The dispersion relations of the modes are thus given by $z_\alpha=z_\alpha^{(0)}+z_\alpha^{(1)}+\cdots$ with $z_\alpha^{(0)}=0$ for $\alpha=7$ and $\alpha=8$.  We emphasize that the dispersion relations depend not only on the magnitude $q=\Vert{\bf q}\Vert$ of the wave vector, but also on its direction ${\bf e}_{\rm l}={\bf q}/q$.


\section{Vacancy correlation and spectral functions}
\label{app:S_vac}

We define the Fourier transform of the microscopic molar fraction of vacancies~(\ref{y-micro})
\be
\hat y({\bf q},t) \equiv  \int_V \hat y({\bf r},t) \, {\rm e}^{{\rm i}{\bf q}\cdot{\bf r}} \, {\rm d}{\bf r} =n_{\rm eq,0}^{-1}  \sum_{i=1}^{N_{\rm v}} {\rm e}^{{\rm i}{\bf q}\cdot{\bf r}_{{\rm v},i}(t)} \, ,
\ee
the correlation function 
\begin{align}
\label{eq:ISFV}
F_{\rm vac}({\bf q},t) &\equiv \frac{1}{N_{\rm v}} \langle \hat{y}({\bf q},t)\, \hat{y}^*({\bf q},0)\rangle_{\rm eq}\notag\\
& = \frac{1}{n_{\rm eq,0}^{2}N_{\rm v}} \left\langle\sum_{i,j=1}^{N_{\rm v}} {\rm e}^{{\rm i} {\bf q}\cdot\left[{\bf r}_{{\rm v},i}(t)-{\bf r}_{{\rm v},j}(0)\right]}\right\rangle_{\rm eq} 
\end{align}
for ${\bf q}\ne 0$, and the associated spectral function
\begin{align}
\label{eq:VSF-dfn}
S_{\rm vac}({\bf q},\omega) \equiv \int_{-\infty}^{+\infty} F_{\rm vac}({\bf q},t) \, {\rm e}^{-{{\rm i}}\omega t} \, {\mathrm d}t \, ,
\end{align}
such that $\int_{-\infty}^{+\infty} S_{\rm vac}({\bf q},\omega) \, {\rm d}\omega/(2\pi)=F_{\rm vac}({\bf q},0)$. Hydrodynamics predicts that the spectral function should behave as the following Lorentzian function,
\begin{align}
\label{eq:VSF-hydro}
S_{\rm vac}({\bf q},\omega) \simeq F_{\rm vac}({\bf q},0) \, \frac{ 2\, D_{\rm vac} q^2}{\omega^2+(D_{\rm vac} q^2)^2} \, ,
\end{align}
where the diffusivity $D_{\rm vac}$ is given by Eqs.~\eqref{D_vac-final-1}-\eqref{D_vac-final-2}, which depends on the direction of the wave vector $\bf q$. 
The spectral function~(\ref{eq:VSF-hydro}) has poles at the complex frequencies $\omega=\pm{\rm i} D_{\rm vac} q^2$, corresponding to the dispersion relation of the eighth mode.


\section{Markovian jump model}
\label{app:Markov}

The simplest model to describe the random walk of a vacancy is the following Markov jump process.  If all the jumps happen between nearest-neighboring sites (thus, having the length $a/\sqrt{2}$) and if the vacancy does not jump before a very large number of binary collisions in the system, we may assume that the events of vacancy jumps are identical and statistically independent.  In this case, we may consider the Markov jump process ruled by the following master equation,
\be
\label{master-eq-1}
\frac{d}{dt} P({\bf r},t) = \frac{w}{12} \sum_{n=1}^{12} \left[ P({\bf r}-\Delta{\bf r}_n,t) - P({\bf r}, t) \right] ,
\ee
where ${\bf r}$ denotes the discrete position of the lattice site where the vacancy is located, $w$ is the transition rate, $\Delta{\bf r}_n$ are the jumps of its position to its twelve nearest neighbors
\be\label{jump_vect}
\Delta{\bf r}_n \in\left\{ \left(0,\pm\frac{a}{2},\pm\frac{a}{2}\right), \left(\pm\frac{a}{2},0,\pm\frac{a}{2}\right), \left(\pm\frac{a}{2},\pm\frac{a}{2},0\right) \right\} .
\ee
For this Markov jump process, the probability distribution of the jump times is the exponential distribution $p(\tau) = w \, {\rm e}^{-w\tau}$.

Introducing the probability density $p({\bf r},t) \equiv a^{-3} P({\bf r},t)$ and expanding in powers of $\Delta{\bf r}_n$ in the scaling limit where $a\Vert\boldsymbol\nabla p\Vert/p \ll 1$, we obtain the diffusion equation $\partial_t\, p = D_{\rm vac}\nabla^2 p$ with the vacancy diffusion coefficient given by Eq.~(\ref{eq:D_omega}).  For this model, diffusion is isotropic and $D_{\rm vac}={\cal D}_y$, since there is a single vacancy, so that $y\simeq 0$.

The solutions of the master equation~(\ref{master-eq-1}) can be decomposed into the Fourier modes $P({\bf r},t)\sim {\rm e}^{-{\rm i}{\bf q}\cdot{\bf r}+z({\bf q})t}$ with the following exact dispersion relation,
\be\label{Markov_disp_rel}
z({\bf q}) = \frac{w}{12} \sum_{n=1}^{12} \left( {\rm e}^{{\rm i}{\bf q}\cdot\Delta{\bf r}_n} - 1 \right) ,
\ee
which defines the effective diffusion coefficient $D_{\rm vac}^{\rm (eff)}({\bf q})\equiv - z({\bf q})/q^2$.  The latter reduces to the expression~(\ref{eq:D_omega}) in the hydrodynamic limit $q\to 0$.  Using the jump vectors~(\ref{jump_vect}), the exact dispersion relation becomes
\begin{align}
z(q_x,q_y,q_z) =&\ \frac{w}{3} \bigg( \cos\frac{aq_x}{2} \, \cos\frac{aq_y}{2} +  \cos\frac{aq_y}{2} \, \cos\frac{aq_z}{2} \nonumber\\
&+ \cos\frac{aq_x}{2} \, \cos\frac{aq_z}{2} - 3  \bigg) .
\label{Markov_disp_rel-fcc}
\end{align}
Since the decay of the Fourier modes of Eq.~(\ref{master-eq-1}) is exponential, the spectral function of this model has the following  Lorentzian form,
\begin{align}
\label{eq:VSF-Markov}
S_{\rm vac}({\bf q},\omega) = F_{\rm vac}({\bf q},0) \, \frac{ - 2\, z({\bf q})}{\omega^2+z({\bf q})^2} \, ,
\end{align}
where $z({\bf q})$ is the dispersion relation~(\ref{Markov_disp_rel-fcc}) and $F_{\rm vac}({\bf q},0)=n_{{\rm eq},0}^{-2}$ because of the definition~(\ref{eq:ISFV}) for $N_{\rm v}=1$.
We note that the hydrodynamic limit is reached if $aq\ll 1$, where we recover the dispersion relation $z({\bf q})= -(w a^2/12)\, q^2 + O(q^4)$ of the diffusive approximation.


\newpage
\pagebreak

\begin{widetext}

\begin{table}[!htbp]
  \begin{tabular}{c @{\hskip .8cm} c @{\hskip 0.5cm} c @{\hskip 0.5cm} c @{\hskip 0.5cm} c  }
    \hline\hline
 		$n_{0*}$ &   $y=1/32$ &  $y=1/108$ &   $y=1/256$ &  $y=1/500$  \\
    \hline  
     	1.05   &  $(3.89\pm0.01)\times 10^{-4}$  & $(1.55\pm0.01)\times 10^{-4}$ & $(0.71\pm0.01)\times 10^{-4}$  & $(0.37\pm0.01)\times 10^{-4}$ \\    
     	1.075	  &  $(1.28\pm0.01)\times 10^{-4}$ & $(0.56\pm0.01)\times 10^{-4}$& $(0.26\pm0.01)\times 10^{-4}$ & $(0.14\pm0.01)\times 10^{-4}$ \\   
     	1.1	  &  $(2.99\pm0.02)\times 10^{-5	}$ & $(1.45\pm0.01)\times 10^{-5}$ & $(0.72\pm0.01)\times 10^{-5}$ & $(0.40\pm0.02)\times 10^{-5}$  \\   
     	1.125	  &  $(4.11\pm0.04)\times 10^{-6	}$   & $(2.53\pm0.02	)\times 10^{-6}$& $(1.27\pm0.02	)\times 10^{-6}$ &  $(0.73\pm0.18	)\times 10^{-6}$  \\ 
    \hline\hline
    \\
  \end{tabular}
  \caption{Vacancy conductivity $\zeta$ versus the density $n_{0}$ and the vacancy molar fraction $y$ computed with the method of the Helfand moments. Parameters used in the simulations are $\Delta t_* = 0.1$, $t_{\rm transient *}=5.0$, $n_{\rm step}=100$, $N_{\rm stat}=10^6$ for $y\in\{1/32,1/108\}$, $N_{\rm stat}=2\times10^5$ for $y=1/256$, and $N_{\rm stat}=10^5$ for $y=1/500$. The number of lattice sites is $N_0=1/y$ and the number of vacancies $N_{\rm v}=1$.}\label{Tab:zeta_H}
\end{table}


\begin{table}[!htbp]
  \begin{tabular}{c @{\hskip .8cm} c @{\hskip 0.5cm} c @{\hskip 0.5cm} c @{\hskip 0.5cm} c    }
    \hline\hline
 		$n_{0*}$ &   $y=1/32$ &   $y=1/108$ &  $y=1/256$ &  $y=1/500$   \\
    \hline  
     	1.05	  &  $-(1.26\pm0.09)\times 10^{-3}$   & $-(0.50\pm0.05)\times 10^{-3}$& $-(0.28\pm0.14)\times 10^{-3}$  & $-(0.21\pm0.15)\times 10^{-3}$ \\    
     	1.075	  &  $-(3.74\pm0.85)\times 10^{-4}$ & $-(1.90\pm0.42)\times 10^{-4}$& $-(0.98\pm0.92)\times 10^{-4}$  & $-(0.68\pm0.44)\times 10^{-4}$  \\   
     	1.1	  &  $-(7.76\pm6.52)\times 10^{-5}$  & $-(4.38\pm1.64)\times 10^{-5}$& $-(2.58\pm4.27)\times 10^{-5}$  & $-(1.68\pm4.84)\times 10^{-5}$   \\   
     	1.125	  &  $-(1.55\pm0.69)\times 10^{-5}$  & $-(0.99\pm0.69)\times 10^{-5}$ & $-(0.50\pm0.97)\times 10^{-5}$  & $-(0.33\pm3.02)\times 10^{-5}$  \\ 

    \hline\hline
     \end{tabular}
  \caption{Coefficient $\xi$ of vacancy thermodiffusion versus  the density $n_{0}$ and the vacancy molar fraction $y$ computed with the method of the Helfand moments. Parameters used in the simulations are $\Delta t_* = 0.1$, $t_{\rm transient *}=5.0$, $n_{\rm step}=100$, $N_{\rm stat}=10^6$ for $y\in\{1/32,1/108\}$, $N_{\rm stat}=2\times10^5$ for $y=1/256$, and $N_{\rm stat}=10^5$ for $y=1/500$. The number of lattice sites is $N_0=1/y$ and the number of vacancies $N_{\rm v}=1$.}\label{Tab:xi_H}
\end{table}

\begin{table}[!htbp]
  \begin{tabular}{c @{\hskip .8cm} c @{\hskip 0.5cm} c @{\hskip 0.5cm} c @{\hskip 0.5cm} c     }
    \hline\hline
 		$n_{0*}$ &   $y=1/32$ &   $y=1/108$ &  $y=1/256$ &  $y=1/500$  \\
    \hline  
     	1.05	  &  $12.01\pm0.02$   & $13.35\pm0.02$ & $13.76\pm 0.03$ & $ 13.95\pm 0.06$\\    
     	1.075	  &  $13.86\pm0.94$ & $14.49\pm0.04$& $14.91\pm 0.07$ & $15.09\pm0.12$ \\   
     	1.1	  &  $14.11\pm0.01$  & $15.85\pm0.02$& $16.45\pm0.08$ & $16.85\pm0.69$  \\  
     	1.125	  &  $15.51\pm0.02$   & $17.52\pm0.01$ & $18.08\pm0.09$ &  $18.49\pm0.58$ \\ 
    \hline\hline
      \end{tabular}
  \caption{Heat conductivity  $\kappa$ versus of the density $n_{0}$ and the vacancy molar fraction $y$ computed with the method of the Helfand moments. Parameters used in the simulations are $\Delta t_* = 0.1$, $t_{\rm transient *}=5.0$, $n_{\rm step}=100$, $N_{\rm stat}=10^6$ for $y\in\{1/32, 1/108\}$, $N_{\rm stat}=2\times10^5$ for $y=1/256$, and $N_{\rm stat}=10^5$ for $y=1/500$. The number of lattice sites is $N_0=1/y$ and the number of vacancies $N_{\rm v}=1$.}\label{Tab:kappa_H}
\end{table}

\begin{table}[!htbp]
  \begin{tabular}{c @{\hskip .8cm} c @{\hskip 0.5cm} c @{\hskip 0.8cm} c @{\hskip 0.5cm} c    @{\hskip 0.5cm}}
    \hline\hline
 		$n_{0*}$ &   $y=1/500$ &   $y=1/1000$ &  $y=1/2000$   \\
    \hline  
     	1.05   &  $(3.73\pm0.03)\times 10^{-5}$  & $(1.97\pm0.03)\times 10^{-5}$ & $(1.00\pm0.05)\times 10^{-5}$ &  \\    
     	1.075	  &  $( 1.38\pm0.02)\times 10^{-5}$  & $(0.88\pm0.04)\times 10^{-5}$ & $(0.35\pm0.03)\times 10^{-5}$  \\   
     	1.1	  &  $(3.99\pm0.12)\times 10^{-6	}$  & $(1.92\pm0.19)\times 10^{-6}$ & $(1.05\pm0.14)\times 10^{-6}$  \\   
     	1.125	  &  $(7.32\pm1.80)\times 10^{-7	}$  & $(4.16\pm1.11)\times 10^{-7}$ & $(2.07\pm0.51)\times 10^{-7}$ \\ 
    \hline\hline
  \end{tabular}
  \caption{Vacancy conductivity $\zeta$ versus the density $n_{0}$ and the vacancy molar fraction $y$ computed with the method of the Helfand moments. Parameters used in the simulations are $\Delta t_* = 0.01$, $t_{\rm transient *}=5$, $n_{\rm step}=100$, $N_{\rm stat}=10^4$ for $y\in\{1/500, 1/1000, 1/2000\}$. The values for $y=1/1000$ have been computed with two vacancies in the dilute limit, the values for $y\in\{1/500,1/2000\}$ with one vacancy. }\label{Tab:zeta_y_H}
\end{table}


\begin{table}[!htbp]
  \begin{tabular}{c @{\hskip .8cm} c @{\hskip 0.5cm} c @{\hskip 0.5cm} c @{\hskip 0.5cm} c  }
    \hline\hline
 		$n_{0*}$ &   $y=1/32$ &   $y=1/108$ &  $y=1/256$ &  $y=1/500$    \\
    \hline  
     	1.05	  &  $(13.1\pm0.1)\times 10^{-3}$   & $(17.6\pm0.1)\times 10^{-3}$  & $(19.0\pm0.2)\times 10^{-3}$ & $(19.6\pm0.2)\times 10^{-3}$ \\    
     	1.075	  &  $(44.1\pm0.2)\times 10^{-4}$  & $(64.8\pm0.2)\times 10^{-4}$& $(71.9\pm0.5)\times 10^{-4}$ & $(74.1\pm1.1)\times 10^{-4}$ \\   
     	1.1	  &  $(10.5\pm0.1)\times 10^{-4}$   & $(17.3\pm0.1)\times 10^{-4}$& $(20.2\pm0.3)\times 10^{-4}$ & $(22.0\pm0.6)\times 10^{-4}$ \\   
     	1.125	  &  $(14.8\pm0.2)\times 10^{-5}$  & $(30.7\pm0.2)\times 10^{-5}$& $(36.6\pm0.6)\times 10^{-5}$ & $(41.2\pm1.0)\times 10^{-5}$  \\ 

    \hline\hline
  \end{tabular}
  \caption{Vacancy diffusion coefficient ${\cal D}_y$ versus the density $n_{0}$ and the vacancy molar fraction $y$. ${\cal D}_y$ is obtained from Eq.~\eqref{D-zeta} using the data for $\zeta$ in Table~\ref{Tab:zeta_H}.}\label{Tab:Dy_H}
\end{table}

\begin{table}[!htbp]
  \begin{tabular}{c @{\hskip .8cm} c @{\hskip 0.5cm} c @{\hskip 0.8cm} c @{\hskip 0.5cm} c    }
    \hline\hline
 		$n_{0*}$ &   $y=1/32$ &   $y=1/108$ &  $y=1/256$ &  $y=1/500$    \\
    \hline  
     	1.05	  &  $(13.2\pm0.2)\times 10^{-3}$  & $(18.3\pm0.3)\times 10^{-3}$ &  $(18.9\pm0.8)\times 10^{-3}$ &  $(21.0\pm2.2)\times 10^{-3}$ \\    
     	1.075	  &  $(44.4\pm0.5)\times 10^{-4}$  & $(65.4\pm1.7)\times 10^{-4}$&  $(74.6\pm2.6)\times 10^{-4}$ &  $(77.3\pm1.2)\times 10^{-4}$ \\   
     	1.1	  &  $(10.5\pm0.2)\times 10^{-4}$ & $(17.5\pm0.8)\times 10^{-4}$&  $(20.7\pm0.4)\times 10^{-4}$ & $(22.4\pm0.6)\times 10^{-4}$ \\   

     	1.125	  &  $(14.6\pm0.9)\times 10^{-5}$   & $(32.5\pm0.1)\times 10^{-5}$ &  $(38.3\pm0.2)\times 10^{-5}$ & $(39.5\pm1.8)\times 10^{-5}$ \\ 

    \hline\hline
  \end{tabular}
  \caption{Vacancy diffusion coefficient ${\cal D}_y$ versus the density $n_{0}$ and the vacancy molar fraction $y$. ${\cal D}_y$ is obtained from Eq.~\eqref{eq:D_omega} based on the jump frequencies for the Markovian jump model of the random walk of a single vacancy in the crystal. Parameters used in the simulations are $\Delta t_* = 0.1$, $t_{\rm transient *}=5.0$, $n_{\rm step}=100$, $N_{\rm stat}=10^6$ for $y\in\{1/32, 1/108\}$, $N_{\rm stat}=2\times10^5$ for $y=1/256$, and $N_{\rm stat}=10^5$ for $y=1/500$. The number of lattice sites is $N_0=1/y$ and the number of vacancies $N_{\rm v}=1$.}\label{Tab:Dy_omega}
\end{table}


\begin{table}[!htbp]
  \begin{tabular}{c @{\hskip .8cm} c @{\hskip 0.5cm} c @{\hskip 0.8cm} c     }
    \hline\hline
 		$n_{0*}$ &   $y=1/32$ &   $y=1/108$ &  $y=1/256$    \\
    \hline  
     	1.05	  &  $(11.5\pm0.1)\times 10^{-3}$ & $(16.2\pm0.1)\times 10^{-3}$ & $(17.7\pm0.1)\times 10^{-3}$  \\    
     	1.075	  &  $(37.5\pm0.1)\times 10^{-4}$   & $(59.0\pm0.1)\times 10^{-4}$& $(71.3\pm0.1)\times 10^{-4}$   \\   
    \hline\hline
      \end{tabular}
  \caption{Diffusivity $D_{\rm vac}$ of the vacancy diffusion mode versus the vacancy molar fraction $y$ for the densities $n_{0*}\in\{1.05,1.075\}$ computed from the spectral functions. The vector ${\bf q}$ is along the $[100]$ direction and has a magnitude $q=2\pi/V^{1/3}$. For $n_{0*}=1.05$, the parameters used in the simulations are $\Delta t_* = 0.2$, $n_{\rm step}=10^3$, $N_{\rm stat}=10^4$.  For $n_{0*}=1.075$,  the parameters are $\Delta t_* = 0.4$, $n_{\rm step}=10^3$, $N_{\rm stat}=10^4$ for $y\in\{1/32,1/108\}$ and $\Delta t_* = 0.5$, $n_{\rm step}=10^3$, $N_{\rm stat}=10^4$ for $y=1/256$. The number of lattice sites is $N_0=1/y$ and the number of vacancies $N_{\rm v}=1$.}\label{Tab:D_vac_SF}
\end{table}


\begin{table}[!htbp]
  \begin{tabular}{c @{\hskip .8cm} c @{\hskip 0.5cm} c @{\hskip 0.5cm} c @{\hskip 0.5cm} c @{\hskip 0.5cm} c @{\hskip 0.5cm} c @{\hskip 0.5cm} c   @{\hskip 0.5cm} c }
    \hline\hline
 		$n_{0*}$ &   $p_*$ &   $\pi_{y\ast}$ &  $B_{T \ast}$ &  $c_{p\ast}$ &  $C_{11\ast}$ &  $C_{12\ast}$ &  $C_{44\ast}$   &     $y_{\rm eq}$ \\
    \hline  
     	1.05	  &  $12.1\pm0.1$  & $-3.2\pm0.3$ & $45.0\pm0.1$ & $4.59\pm0.01$ & $79.0\pm0.3$  & $21.8\pm0.2$ & $ 48.5\pm2.5$& $(17.0\pm0.8)\times10^{-5}$\\    
     	1.075	  &  $ 13.2\pm0.1$  & $-6.0\pm0.4$ & $53.9\pm0.1$ & $4.53\pm0.01$  & $95.2\pm 0.5$ & $26.5\pm0.3$ & $58.2\pm2.3$ & $(5.4\pm0.5)\times10^{-5}$\\   
     	1.1	  &  $14.6\pm0.1$  & $-8.9\pm0.3$ & $64.9\pm0.1$ & $4.48\pm0.01$ & $117.0\pm 0.6$ & $31.5\pm0.3$ & $70.8\pm 1.7$& $(1.5\pm0.3)\times10^{-5}$\\   
     	1.125	  &  $16.2\pm0.1$  & $-11.6\pm0.5$ & $78.9\pm0.1$ & $4.46\pm0.01$ & $143.5\pm 0.5$ & $38.5\pm 0.2$ & $88.0\pm3.5$& $(3.8\pm0.8)\times10^{-6}$ \\ 
    \hline\hline
  \end{tabular}
  \caption{Thermodynamic and elastic properties versus the density $n_{0}$. $p$ is the pressure, $\pi_y$ the derivative of the pressure with respect to $y$, $B_T$ the isothermal bulk modulus, $c_p$ the specific heat capacity, $C_{11}$, $C_{12}$, and $C_{44}$ the three elastic constants of a cubic crystal, and $y_{\rm eq}$ the equilibrium density of vacancies given in Ref.~\cite{L20}.}\label{Tab:TH_EC}
\end{table}


\pagebreak
\end{widetext}




\begin{thebibliography}{99}

\bibitem{N60} Y.~Nambu, {\it Quasiparticles and gauge invariance in the theory of
  superconductivity}, Phys. Rev. {\bf 117}, 648-663 (1960).

\bibitem{G61} J.~Goldstone, {\it Field theories with superconductor solutions}, Il Nuovo Cimento (1955-1965) {\bf 19}, 154-164 (1961).

\bibitem{F75} D. Forster, {\it Hydrodynamic Fluctuations, Broken Symmetry, and Correlation Functions} (Benjamin/Cummings, Reading MA, 1975).

\bibitem{MPP72} P.~C. Martin, O.~Parodi, and P.~S. Pershan, {\it Unified hydrodynamic theory for crystals, liquid crystals, and normal fluids}, Phys. Rev. A {\bf 6}, 2401-2420 (1972).

\bibitem{FC76} P.~D. Fleming and C.~Cohen, {\it Hydrodynamics of solids}, Phys. Rev. B {\bf 13}, 500-516 (1976).

\bibitem{KDEP90} T. R. Kirkpatrick, S. P. Das, M. H. Ernst, and J. Piasecki, {\it Kinetic theory of transport in a hard sphere crystal}, J. Chem. Phys. {\bf 92}, 3768-3780 (1990).

\bibitem{MG24a} J. Mabillard and P.~Gaspard, {\it Hydrodynamic properties of the perfect hard-sphere crystal: Microscopic computations with Helfand moments}, J. Stat. Mech.: Theory Exp. {\bf 2024}, 023208 (2024).

\bibitem{MG24b} J. Mabillard and P.~Gaspard, {\it Hydrodynamic correlation and spectral functions of perfect cubic crystals}, J. Stat. Mech.: Theory Exp. {\bf 2024}, 033204 (2024).

\bibitem{MG24c} J. Mabillard and P.~Gaspard, {\it Elastic and transport coefficients of the perfect hard-sphere crystal from the poles of the hydrodynamic spectral functions}, J. Stat. Mech.: Theory Exp. {\bf 2024}, 033205 (2024).

\bibitem{PBHB24} S. Pieprzyk, A. C. Bra\'nka, D. M. Heyes, and M. N. Bannerman, {\it Revised Enskog theory and molecular dynamics simulations of the viscosities and thermal conductivity of the hard-sphere fluid and crystal}, Phys. Rev. E {\bf 109}, 054119 (2024).

\bibitem{HL64} R.~E. Howard and A.~B. Lidiard, {\it Matter transport in solids}, Rep. Prog. Phys. {\bf 27}, 161-240 (1964).

\bibitem{AL87} A.~R. Allnatt and A.~B. Lidiard, {\it Statistical theories of atomic transport in crystalline solids}, Rep. Prog. Phys. {\bf 50}, 373-472 (1987).

\bibitem{PF01} S. Pronk and D. Frenkel, {\it Point Defects in Hard-Sphere Crystals}, J. Phys. Chem. B {\bf 105}, 6722-6727 (2001).

\bibitem{OGHLRS10} M. Oettel, S. G\"orig, A. H\"artel, H. L\"owen, M. Radu, and T. Schilling, {\it Free energies, vacancy concentrations, and density distribution anisotropies in hard-sphere crystals: {A} combined density functional and simulation study}, Phys. Rev. E {\bf 82}, 051404 (2010).

\bibitem{FGHNKJV14} C. Freysoldt, B. Grabowski, T. Hickel, J. Neugebauer, G. Kresse, A. Janotti, and C. {Van de Walle}, {\it First-principles calculations for point defects in solids}, Rev. Mod. Phys. {\bf 86}, 253-305 (2014).

\bibitem{L20} J. F. Lutsko, {\it Explicitly stable fundamental-measure-theory models for classical density functional theory}, Phys. Rev. E {\bf 102}, 062137 (2020).

\bibitem{LC78} F. C. Larch\'e and J. W. Cahn, {\it Thermochemical equilibrium of multiphase solids under stress}, Acta Metall. {\bf 26}, 1579-1589 (1978).

\bibitem{LC85} F. C. Larch\'e and J. W. Cahn, {\it The interactions of composition and stress in crystalline solids}, Acta Metall. {\bf 33}, 331-357 (1985).

\bibitem{CVJ18} E. Clouet, C. Varvenne, and T. Jourdan, {\it Elastic modeling of point defects and their interaction}, Computational Material Science {\bf 147}, 49-63 (2018).

\bibitem{CL83} J. W. Cahn and F. C. Larch\'e, {\it An invariant formulation of multicomponent diffusion in crystals}, Scripta Metallurgica {\bf 17}, 927-932 (1983).

\bibitem{KBR89} K. W. Kehr, K. Binder, and S. M. Reulein, {\it Mobility, interdiffusion, and tracer diffusion in lattice-gas models of two-component alloys}, Phys. Rev. B {\bf 39}, 4891-4910 (1989).

\bibitem{GM83} H. Grabert and W. Michel, {\it Nonlinear transport equations and invariance principles for solids}, Phys. Lett. A {\bf 98}, 183-186 (1983).

\bibitem{S97} G.~Szamel, {\it Statistical mechanics of dissipative transport in crystals}, J. Stat. Phys. {\bf 87}, 1067-1082 (1997).

\bibitem{MG20} J. Mabillard and P.~Gaspard, {\it Microscopic approach to the macrodynamics of matter with broken symmetries}, J. Stat. Mech.: Theory Exp. {\bf 2020}, 103203 (2020).

\bibitem{MG21} J. Mabillard and P. Gaspard, {\it Nonequilibrium statistical mechanics of crystals}, J. Stat. Mech.: Theory Exp. {\bf 2021}, 063207 (2021).

\bibitem{H22} R. Haussmann, {\it Microscopic density-functional approach to nonlinear elasticity theory}, J. Stat. Mech.: Theory Exp. {\bf 2022}, 053210 (2022).

\bibitem{H23} K. Hiura, {\it Microscopic derivation of nonlinear fluctuating hydrodynamics for crystalline solid}, Phys. Rev. E {\bf 108}, 054101 (2023).

\bibitem{GM84} S.~R.~de Groot and P.~Mazur, {\it Nonequilibrium Thermodynamics} (Dover, New York, 1984).

\bibitem{G52}  M. S. Green, {\it Markoff Random Processes and the Statistical Mechanics of Time-Dependent Phenomena}, J. Chem. Phys. {\bf 20}, 1281-1295 (1952).

\bibitem{G54}  M. S. Green, {\it Markoff Random Processes and the Statistical Mechanics of Time-Dependent Phenomena. {II.}~{Irreversible} Processes in Fluids}, J. Chem. Phys. {\bf 22}, 398-413 (1954).

\bibitem{K57} R. Kubo, {\it Statistical mechanical theory of irreversible processes. {I.} {General} theory and simple applications in magnetic and conduction problems}, J. Phys. Soc. Japan {\bf 12}, 570-586 (1957).

\bibitem{M58} H. Mori, {\it Statistical-Mechanical Theory of Transport in Fluids}, Phys. Rev. {\bf 112}, 1829-1842 (1958).

\bibitem{McL63} J.~A. McLennan~Jr., {\it The formal statistical theory of transport processes}, Adv. Chem. Phys. {\bf 5}, 261-317 (1963).

\bibitem{R66} B. Robertson, {\it Equations of Motion in Nonequilibrium Statistical Mechanics}, Phys. Rev. {\bf 144}, 151-161 (1966).

\bibitem{P68} R. A. Piccirelli, {\it Theory of the dynamics of simple fluids for large spatial gradients and long memory}, Phys. Rev. {\bf 175},  77-98 (1968).

\bibitem{OL79}
I.~Oppenheim and R.~D. Levine, {\it Nonlinear transport processes: Hydrodynamics}, Physica A {\bf 99}, 383-402 (1979).

\bibitem{S14}
S.-i. Sasa, {\it Derivation of hydrodynamics from the Hamiltonian description of particle systems}, Phys. Rev. Lett. {\bf 112}, 100602 (2014).

\bibitem{MG23} J. Mabillard and P.~Gaspard, {\it Quantum local-equilibrium approach to dissipative hydrodynamics}, Phys. Rev. E {\bf 107}, 014102 (2023).

\bibitem{E26} A. Einstein, {\it Investigations on the Theory of the {Brownian} Movement} (E. P. Dutton \& Company, New York, 1926).

\bibitem{H60} E. Helfand, {\it Transport Coefficients from Dissipation in a Canonical Ensemble}, Phys. Rev. {\bf 119}, 1-9 (1960).

\bibitem{W98} D.~C. Wallace, {\it Thermodynamics of Crystals} (Dover, New York, 1998).

\bibitem{BP76}  B. J. Berne and R. Pecora, {\it Dynamic Light Scattering} (Wiley, New York, 1976).
  
\bibitem{BY80}  J. P. Boon and S. Yip, {\it Molecular Hydrodynamics} (McGraw-Hill, New York, 1980).

\bibitem{DvBK21} J. R. Dorfman, H. {van Beijeren}, and T. R. Kirkpatrick, {\it Contemporary Kinetic Theory of Matter} (Cambridge University Press, Cambridge UK, 2021).

\bibitem{H97} J. M. Haile,  {\it Molecular Dynamics Simulation: Elementary Methods} (Wiley, New York, 1997).

\bibitem{AM76} N. W. Ashcroft and N. D. Mermin, {\it Solid State Physics} (HRW International Editions, Philadelphia, PA, 1976).

\bibitem{S98} R. J. Speedy, {\it Pressure and entropy of hard-sphere crystals}, J. Phys.: Condens. Matter {\bf 10}, 4387-4391 (1998).

\bibitem{PBHB20} S. Pieprzyk, A. C. Bra\'nka, D. M. Heyes, and M. N. Bannerman, {\it A comprehensive study of the thermal conductivity of the hard sphere fluid and solid by molecular dynamics simulation}, Phys. Chem. Chem. Phys. {\bf 22}, 8834-8845 (2020).

\bibitem{GAW70} D. M. Gass, B. J. Alder, and T. E. Wainwright, {\it The thermal conductivity of a hard sphere solid}, J. Phys. Chem. Solid {\bf 32}, 1797-1800 (1970).

\bibitem{BA71} C. H. Bennett and B. J. Alder, {\it Studies in Molecular Dynamics. {IX.} {Vacancies} in Hard Sphere Crystals}, J. Chem. Phys. {\bf 54}, 4796-4808 (1971).

\bibitem{vdMDF17} B. van~der~Meer, M. Dijkstra, and L. Filion, {\it Diffusion and interactions of point defects in hard-sphere crystals}, J. Chem. Phys. {\bf 146}, 244905 (2017).

\end{thebibliography}
\end{document}